\documentclass[draf]{aa}
\usepackage{natbib}
\usepackage{color}
\usepackage{graphicx}
\usepackage[varg]{txfonts}
\usepackage{hyperref}
\hypersetup{citecolor=black, urlcolor=blue, linkcolor=black, colorlinks=true}
\newcommand{\sect}[1]{Sect.\,\ref{#1}}

\newcommand{\fig}[1]{Fig.\,\ref{#1}}
\newcommand{\figs}[1]{Figs.\,\ref{#1}}
\newcommand{\tab}[1]{Table\,\ref{#1}}

\definecolor{orange}{rgb}{1,0.4,0.}
\graphicspath{{./}{figs-ps/}{figs/}}
\begin{document}

%=============================================================================
% TITLE
%=============================================================================
%
\title{Investigating explosive events in a 3D quiet-Sun model: Transition region and coronal response} 
\titlerunning{EE and campfire}

%\authorrunning{}

\author{Yajie Chen \inst{1}
          \and
          Hardi Peter \inst{1,2}
          \and
          Damien Przybylski \inst{1}
          }

\institute{
             Max-Planck Institute for Solar System Research, 37077 G\"{o}ttingen, Germany
             \email{cheny@mps.mpg.de}
             \and
             Institut für Sonnenphysik (KIS), 79110 Freiburg, Germany
          }

\date{Version: \today}
%\date{Received xx Jan 2021 / Accepted xx yyy 2021}

  \abstract
  {Transition region explosive events are characterized by non-Gaussian profiles of the emission lines formed at transition region temperatures, 
  and they are believed to be manifestations of small-scale reconnection events in the transition region.
  }
  {
  Traditionally, the enhanced emission at the line wings is interpreted as bi-directional outflows generated by reconnection of oppositely directed magnetic field. 
  We investigate whether the 2D picture also holds in a more realistic setup of a 3D radiation magnetohydrodynamics (MHD) quiet-Sun model.
  We also compare the thermal responses in the transition region and corona of different events.
  }
  {
  We took a 3D self-consistent quiet-Sun model extending from the upper convection zone to the lower corona calculated using the MURaM code.
  We first synthesized the Si~{\sc{iv}} line profiles from the model and then located the profiles which show signatures of bi-directional flows.
  {These tend to appear along network lanes, and most do not reach coronal temperatures.
  We isolated two hot (around 1 MK) events and one cool (order of 0.1 MK) event 
  and examined the magnetic field evolution in and around these selected events.
  Furthermore, we investigated why some explosive events reach coronal temperatures while most remain cool.
  }
  We also examined the emission of these events as seen in the 174 {\AA} passband of Extreme Ultraviolet Imager (EUI) onboard Solar Orbiter and all coronal passbands of Atmospheric Imaging Assembly (AIA) onboard Solar Dynamics Observatory (SDO).
  }
  {
  The field lines around two events reconnect at small angles, i.e., they undergo component reconnection.
  The third case is associated with the relaxation of a highly twisted flux rope.
  All of the three events reveal signatures in the synthesized EUI 174 {\AA} images.
  The intensity variations in two events are dominated by variations of the coronal emissions,
  while the cool component seen in the respective channel contributes significantly to the intensity variation in one case.
  Comparing to the cool event, one hot event is embedded in regions with higher magnetic field strength and heating rates while the densities are comparable, and the other hot event is heated to coronal temperatures mainly because of the low density.
  }
  {
  Small-scale heating events seen in EUV channels of AIA or EUI might be hot or cool. Our results imply that the major difference between the events in which coronal counterparts dominate or not is the amount of converted magnetic energy and/or density in and around the reconnection region.
  }

\keywords{Sun: magnetic fields
      --- Sun: corona
      --- Sun: transition region
      --- Magnetohydrodynamics (MHD)} 
%
%----------------------------------------------------------------------------

\maketitle

%==============================================================================
\section{Introduction\label{S:intro}}
%==============================================================================

Transition region explosive events are small-scale dynamic events identified from  spectral observations of the solar transition region lines.
These events were first reported by \citet{Brueckner1983} and named "turbulent events".
Explosive events are characterized by non-Gaussian line profiles with wing enhancement.
These events are often found at the edge of the chromospheric network \citep[e.g.,][]{Dere1994AdSpR,Innes1997SoPh}.
Some explosive events are associated with flux emergence and cancellation, which are often believed to be the signatures of magnetic reconnection \citep[e.g.,][]{Dere1991JGR,Chae1998},
and the emission enhancements at the line wings is interpreted as bi-directional reconnection outflows \citep{Innes1997}.
The reconnection outflows can sometimes appear as network jets \citep{Tian2014Sci,Chen2019} or nanojets \citep{Chen2017,Chen2020,Antolin2021} in the transition region images taken by the Interface Region Imaging Spectrograph \citep[IRIS,][]{IRIS}.

Many magnetohydrodynamics (MHD) models, mainly in 2D, of magnetic reconnection are constructed to reproduce the line profiles with signatures of bi-directional outflows
\citep[e.g.,][]{Innes1999,Roussev2001c,Innes2015,Ni2021} 
and to understand the temporal evolution and detailed structures during the reconnection processes \citep[e.g.,][]{Peter2019}.
Recently, the Si~{\sc{iv}} line profiles with extreme line broadening or signatures of bi-directional reconnection outflows have also been reproduced in some self-consistent 3D MHD models \citep{Hansteen2017,Hansteen2019}.

In addition to explosive events, extreme-ultraviolet (EUV) brightenings in the quiet-Sun regions have been extensively studied based on narrowband coronal imaging observations for decades \citep[e.g.,][]{Berghmans1998,Aschwanden2000,Madjarska2019,Chitta2021}.
The thermal energy of these events follows a power-law-like distribution \citep{Aschwanden2000,Purkhart2022}.
Recently, the Extreme Ultraviolet Imager \citep[EUI;][]{EUI} onboard Solar Orbiter \citep{SolO} achieved an unprecedented spatial resolution of $\sim$200 km during its perihelion and is able to resolve the smallest EUV brightenings, sometimes named "campfires", which could be observed by far.
\citet{Berghmans2021} found that these events follow a power-law-like distribution and are an extension of previously reported EUV brightenings to a smaller scale.
%
%They also suggested that these events are heated to 10$^{6.1}$ K at heights of 1--5 Mm above the photosphere. 
%
Furthermore, \citet{Panesar2021} and \citet{Kahil2022} found that some of these events are associated with flux emergence and/or cancellation in the photosphere, 
inferring that the small-scale EUV brightenings may be triggered by magnetic reconnection.

The magnetic nature of the small-scale EUV brightenings has been investigated through forward modeling with 3D MHD models.
\citet{Chen2021} constructed a 3D MHD model representing the very quiet Sun and found that the small-scale EUV brightenings in the model are generated through component reconnection between bundles of field lines or untwisting of flux ropes.
Similar magnetic setups of component reconnection are also found in the transition region loop-like brightenings \citep{Skan2023}.
\citet{Tiwari2022} studied small-scale EUV brightenings in a model of an ephemeral emerging flux region with higher levels of magnetic activities in the photosphere
and found that some are related to flux cancellation.

%\textcolor{blue}{However, the article deviates from its stated aims, with a lot of focus on coronal signatures that are rarely seen in explosive events. It is also questionable whether the three events studied are analogous to explosive events. ... As highlighted in the authors’ introduction, previous studies of explosive events have demonstrated that generally they do not have a coronal signature. Yet all three of the events do have a coronal signature, with events 2 and 3 having strong signatures. In particular, event 2 is actually a coronal brightening that has a surrounding transition region (\fig{cut_cf}). I would suggest that events 2 and 3 are not explosive events.}

{The thermal properties and the relationship between explosive events and small-scale EUV brightenings are still elusive.
Explosive events are frequently observed in the transition region lines including the Si~{\sc{iv}}, C~{\sc{iv}}, and O~{\sc{vi}} lines, but they rarely exhibit counterparts in the coronal lines such as the weak Mg~{\sc{x}} line \citep{Teriaca2002}.
However, \citet{Innes2001} reported that several explosive events in the quiet Sun region appear as EUV brightenings in the Transition Region and Coronal Explorer \citep[TRACE,][]{TRACE} 171 {\AA} images.
\citet{Winebarger2002} found that 35\% of explosive events in an active region exhibit significant intensity variations in the TRACE 171 {\AA} passband, particularly one event that exhibits strong emission in the Ne~{\sc{viii}} line (see Fig. 1 therein).
Two possible explanations for the intensity enhancement in TRACE images are that some explosive events are heated to coronal temperatures, or the emission enhancement of the transition region lines within the spectral window leads to the brightenings.
Thus, while explosive events may be dominated by cool plasma at transition region temperatures, it is unclear whether all of them are not heated to coronal temperatures.
}

Differential emission measurements (DEM) inversion of coordinated Atmospheric Imaging Assembly \citep[AIA,][]{AIA} observations implies that small-scale EUV brightenings detected by EUI can reach 1.2 MK \citep{Berghmans2021}, 
while \citet{Dolliou2023} argued that the emission may be dominated by the cool component considering the irresolvable time delay of intensity variation in different AIA coronal passbands 
\citep{Winebarger2013}.
Recently, \citet{Huang2023} examined the spectral profiles taken by Spectral Imaging of the Coronal Environment \citep[SPICE,][]{SPICE} onboard Solar Orbiter within three EUV brightenings observed by EUI and {found that all events exhibit intensity enhancements in the O~{\sc{vi}} lines, while two show response in the Ne~{\sc{viii}} line.
Their findings suggest that not all small-scale EUV brightenings reach coronal temperatures.
\citet{2024A&A...688A..77D} investigated more EUV brightenings and obtained similar results, i.e., all events show strong emission in the transition region lines, but only some are detectable in the Ne~{\sc{viii}} line.
This implies that small-scale EUV brightenings are likely dominated by plasma at transition region temperatures and are not always heated to coronal temperatures.
However, due to the limited spectral resolution of SPICE observations, it remains unclear whether the events studied by \citet{Huang2023} and \citet{2024A&A...688A..77D} are classified as explosive events.
}

{
}
In this paper, we first identified explosive events from a 3D self-consistent quiet-Sun model and selected three examples for a detailed study.
{We investigated the magnetic field structure in and around the explosive events, then examined their thermal properties and responses in the synthesized coronal images.
Furthermore, we investigated why some explosive events are heated to coronal temperatures while most are not, assuming that the hotter events could resemble some of the explosive events in the real Sun.
Our study suggests that events reaching coronal temperatures result from a higher amount of converted magnetic energy and/or lower density compared to those dominated by the cool component.}

%==============================================================================
\section{MHD model and methods \label{S:model}}
%==============================================================================

We took the same quiet-Sun MHD model as the one used in \citet{Chen2021}.
The model is calculated from the coronal extension of the MURaM code \citep{MURaM,MURaM2017}.
{The MURaM code incorporates a numerical diffusion scheme, which acts to smooth out strong gradients in the simulation.
The change in energy due to the reconfiguration of the field is added to the internal energy as numerical resistive heating, see \citet{MURaM2017} for more details.
This allows the code to model reconnection processes, although the scales which can be resolved are limited by the numerical scheme and resolution.
}
The model extends from $\sim$20 Mm below to $\sim$17.5 Mm above the photosphere in the vertical direction with a grid size of 25 km and covers a region of 50$\times$50 Mm$^2$ in the horizontal direction with a grid spacing of $\sim$48.8 km.
The magnetic field in the model is generated by the small-scale dynamo, and the corona self-consistently maintains an average temperature of $\sim$1 MK.
The snapshots have a duration of 16 minutes and a cadence of 20 seconds.
For more details on the MHD model, we refer the reader to \citet{MURaM2017} and \citet{Chen2021}.

%>>>>>>>>>>>>>>>>>>>>>>>>>>>>>>>>>>>>>>>>>>>>>>>>>>>>>>>>>>>>>>>>>>>>>>>>>>>>>>
\begin{table}[]
\centering
\caption[Emission lines used in this study]{Emission lines used in this study}
\begin{tabular}{c|c|c}
\hline
\hline
Ion name & Wavelength [{\AA}] & log $T_{\rm{max}}$\,[K]    \\ \hline
%log T at maximum ionisation ratio
Si {\sc{iv}}    & 1393.755       & 4.90      \\
O {\sc{vi}}     & 173.079        & 5.50      \\
Fe {\sc{x}}     & 174.531        & 6.00      \\ \hline
\end{tabular}\label{tab:lines}
\end{table}
%<<<<<<<<<<<<<<<<<<<<<<<<<<<<<<<<<<<<<<<<<<<<<<<<<<<<<<<<<<<<<<<<<<<<<<<<<<<<<<

In this study, we used the Si~{\sc{iv}} 1394 {\AA} line profiles to identify explosive events in our model.
We first synthesized the emission and profiles of the Si~{\sc{iv}} line following the procedures described in \citet{Chen2022}.
The emissivity at each grid point is calculated following $n_e^2G(n_e,T)$, where $n_e$ and $T$ are electron number density and temperature, respectively.
$G$ is the contribution function as a function of temperature and electron number density given by the CHIANTI atomic database \citep[version 10.0;][]{CHIANTI,CHIANTI10}.
Then we assumed that the line profile at each grid point is a Gaussian with line center given by the velocity along the line of sight and width given by the thermal width.
In this study, we took the line of sight along the vertical direction,
i.e., we assume the model is located at the disk center.
Thus, we integrated the emissivity and line profile along the vertical direction and obtained spatial maps of emission and line profiles.
Furthermore, we degraded the spatial resolution of the Si~{\sc{iv}} line profiles.
To do this, we first convolved the maps at each wavelength position of the Si~{\sc{iv}} line with a 2D Gaussian function, whose kernel has a full width half maximum of 480 km.
Then we rebinned the spatial maps to a pixel size of 240 km, which corresponds to the typical spatial resolution of IRIS.

We applied the k-means clustering algorithm \citep{kmeans} to the degraded Si~{\sc{iv}} 1394 {\AA} line profiles.
Similar to \citet{Bose2019} and \citet{Joshi2022}, we first calculated the total inertia ($\sigma_k$) as a function of the cluster number ($k$).
Then we chose $k=140$ when $\sigma_k$ almost decreases linearly with $k$.
By examining 140 representative profiles for different groups, we isolated those that reveal bi-directional outflows,
i.e., profiles exhibit both blue and red wing enhancement.
Then the profiles belonging to these representative profiles are marked as signatures of explosive events.
Among $\sim$2 million original line profiles, 1.3\% are identified.
%
%we selected three representative examples to perform a case study.

In order to examine the response of these events in narrowband coronal images, we also synthesized the images at EUI 174 {\AA} and AIA 171, 193, 211, 335, 131, and 94 {\AA} passbands following \citet{Chen2021}.
The radiative loss at each grid point is given by $n_e^2R(n_e,T)$, where $R$ is the response function of different passbands that mainly depends on the temperature.
Then we integrated the emissivity for each passband along the vertical direction to obtain the synthesized 2D images.

As the hot and cool components of the EUI 174 {\AA} passband are dominated by Fe~{\sc{x}} 174 and O~{\sc{vi}} 173 {\AA} lines, respectively,
we further synthesized the emission of the Fe~{\sc{x}} and O~{\sc{vi}} lines following the procedures for Si~{\sc{iv}} emission calculation.
The lines used in this study are listed in \tab{tab:lines}.
Nevertheless, there are more than ten lines except for the strongest Fe~{\sc{x}} 174 and O~{\sc{vi}} 173 {\AA} lines within the spectral window of the EUI 174 {\AA} passband, the quantitative analyses of cool and hot components would require the calculation of intensities of much more spectral lines.
Alternatively, we defined two response functions of $R_c$ and $R_h$:
\begin{equation*}
\begin{split}
    R_c(n_e,T)=\left \{
    \begin{array}{ll}
        R_{174}(n_e,T), & \mathrm{log}_{10}T/K\leqslant 5.75 \\
        0, & \mathrm{log}_{10}T/K\geqslant 5.75
    \end{array}
    \right.
\end{split}
\end{equation*}
\begin{equation*}
\begin{split}
    R_h(n_e,T)=\left \{
    \begin{array}{ll}
        0, & \mathrm{log}_{10}T/K\leqslant 5.75 \\
        R_{174}(n_e,T), & \mathrm{log}_{10}T/K\geqslant 5.75
    \end{array}
    \right.
\end{split}
\end{equation*}
where $R_{174}$ is the response function of EUI 174 {\AA} passband.
Then we recalculated the emissivity following $n_e^2R_c$ and $n_e^2R_h$ and integrated them along the vertical direction, respectively.
In this way, we divided the EUI 174 {\AA} intensity ($I_{174}$) into two parts for simplification: hot component contributed from plasma with temperatures above 10$^{5.75}$ K ($I_{174}^{h}$) and cool component related to plasma with temperatures below 10$^{5.75}$ K ($I_{174}^{c}$),
and $I_{174}$ is essentially equal to the sum of $I_{174}^{h}$ and $I_{174}^{c}$.
%

%>>>>>>>>>>>>>>>>>>>>>>>>>>>>>>>>>>>>>>>>>>>>>>>>>>>>>>>>>>>>>>>>>>>>>>>>>>>>>>
\begin{figure*}
\centering
{\includegraphics[width=16cm]{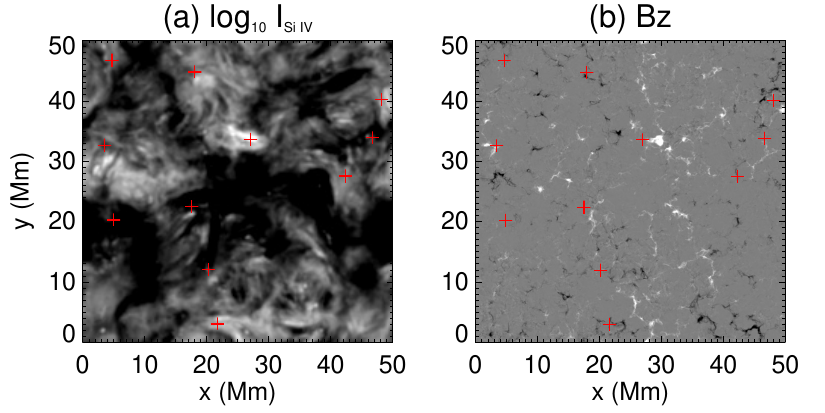}}
\caption{
{(a) Intensity map of the Si\,{\sc{iv}} line integrated over 16 minutes.
{
(b) Magnetogram averaged over 16 minutes, saturated at $\pm$1000 G.}
Red crosses in (a) and (b) indicate the locations of the selected events.
See \sect{S:model}.}
}\label{network}
\end{figure*}
%<<<<<<<<<<<<<<<<<<<<<<<<<<<<<<<<<<<<<<<<<<<<<<<<<<<<<<<<<<<<<<<<<<<<<<<<<<<<<<

%>>>>>>>>>>>>>>>>>>>>>>>>>>>>>>>>>>>>>>>>>>>>>>>>>>>>>>>>>>>>>>>>>>>>>>>>>>>>>>
\begin{figure}
\centering
\includegraphics[width=8cm]{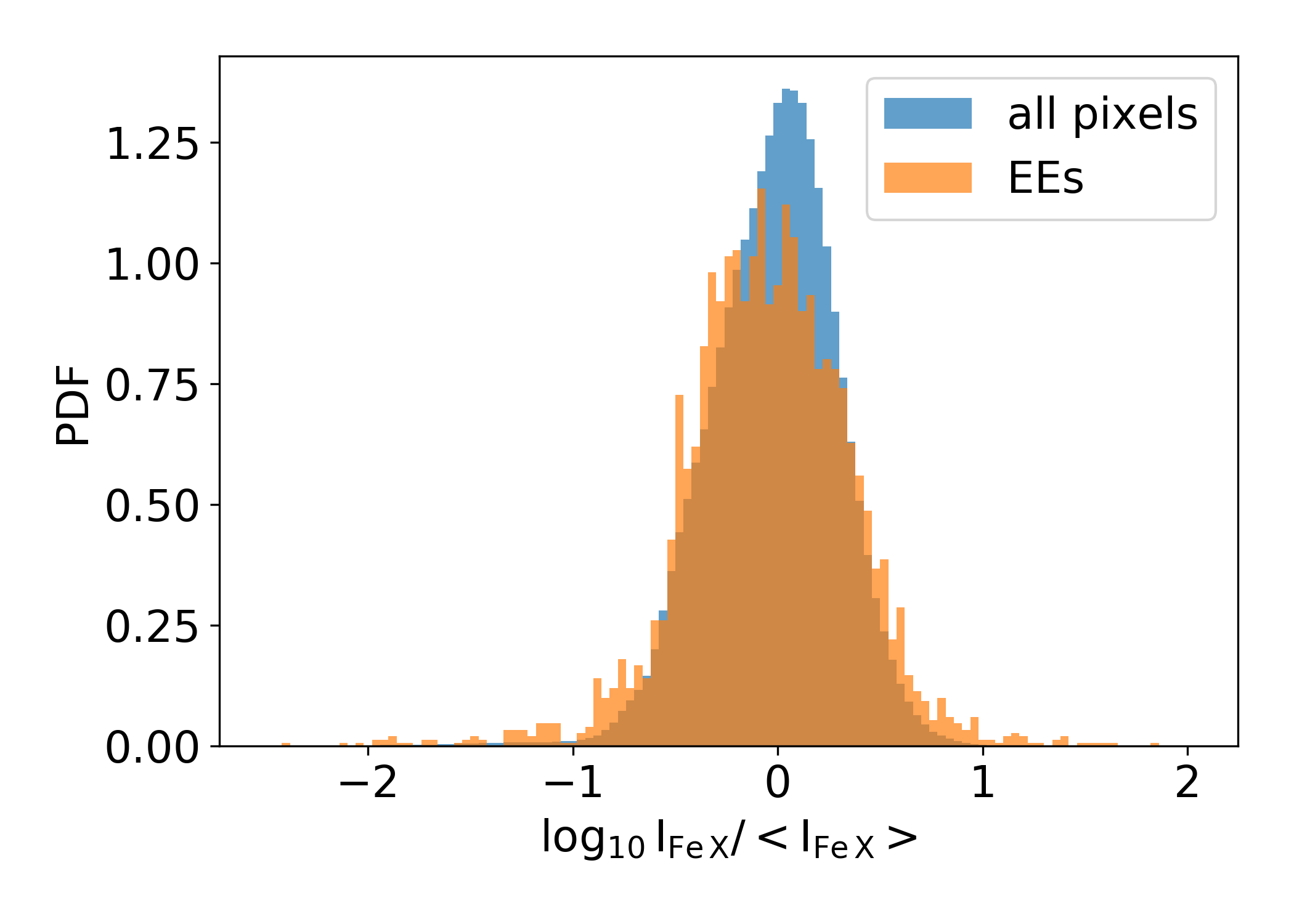}
\caption{
{Probability density function (PDF) of the Fe~{\sc{x}} line intensity for all pixels (in blue) and those identified as explosive events (in orange).
<I$_{Fe~X}$> represents the average intensity of the Fe~{\sc{x}} line across all pixels.
See \sect{S:model}.
}
}\label{intensity_hist}
\end{figure}
%<<<<<<<<<<<<<<<<<<<<<<<<<<<<<<<<<<<<<<<<<<<<<<<<<<<<<<<<<<<<<<<<<<<<<<<<<<<<<<

{We randomly selected 10 events and examined the vertical cuts of temperature and density through these events.
To compare the properties of these events to those of the campfires, we include one event studied in \citet{Chen2021}.
%
%The locations of these events are shown in \fig{network}, mostly along the network lanes.
%
Previous studies have found that explosive events typically occur along network lanes \citep[e.g.,][]{Dere1989,Chae1998}.
The computational domain of our model is sufficiently large to produce several network patches \citep{Chen2021}. We averaged the intensity maps of the Si~{\sc{iv}} line across different snapshots, and the resulting average intensity map is shown in \fig{network}(a). 
{The intensity map presents network patterns, which are associated with small-scale magnetic concentrations where field strengths exceed 1000 G (see \fig{network}(b)).}
Then we marked the locations of the selected explosive events, clearly showing that they are mostly located along the network patterns, in alignment with the previous studies.
{The additional event is the strongest brightening in the Si~{\sc{iv}} line intensity maps, and the Si~{\sc{iv}} line profiles within the event are also identified as explosive events.}
We categorized these events into three groups: those with temperature below 1 MK; those with temperature above 1 MK and high density; those with temperature above 1 MK but low density.
We selected one event from each category and performed detailed analyses for these three events in \sect{S:results}.
Although two of the three events analyzed in this study have obvious coronal counterparts, the Fe~{\sc{x}} line intensity does not show a statistically significant enhancement at the regions where the Si~{\sc{iv}} line profiles are highly deviated from Gaussian {(as shown in \fig{intensity_hist})}.
In other words, most explosive events in our model do not correspond to an enhanced coronal line emission, which is consistent with the previous spectral observations \citep{Teriaca2002}.
}

%==============================================================================
\section{Results \label{S:results}}
%==============================================================================

%------------------------------------------------------------------------------
\subsection{Overview of the three examples}
\label{sp_si4}
%------------------------------------------------------------------------------

%>>>>>>>>>>>>>>>>>>>>>>>>>>>>>>>>>>>>>>>>>>>>>>>>>>>>>>>>>>>>>>>>>>>>>>>>>>>>>>
\begin{figure*}[ht]
\centering {\includegraphics[width=15cm]{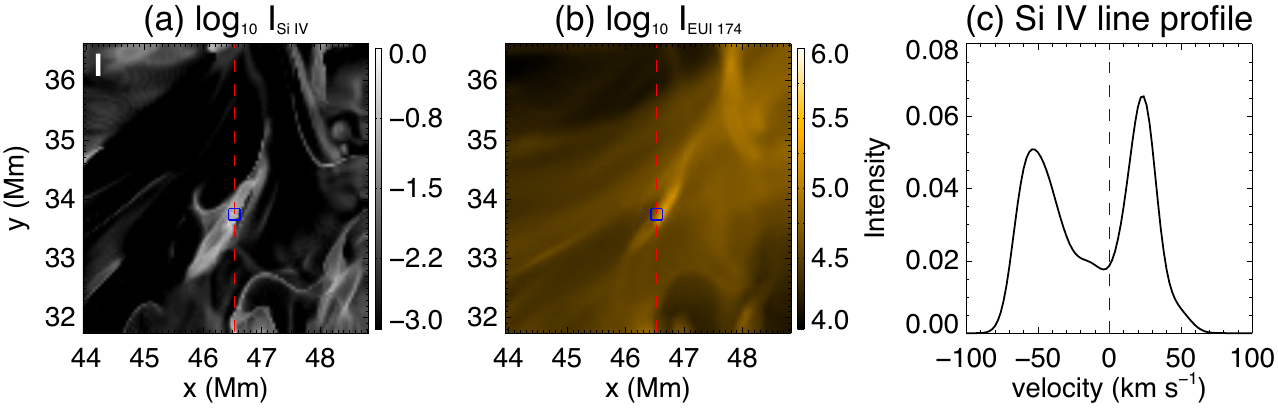}} 
\caption{
Intensity maps and the Si~{\sc{iv}} line profile for Event 1.
(a) Intensity map of the Si~{\sc{iv}} 1394 {\AA} line.
(b) Synthesized emission in EUI 174 {\AA} passband.
(c) The degraded Si~{\sc{iv}} line profile at the blue box shown in panels (a--b). The size of the blue box is $\sim$240 km.
The intensity is shown in arbitrary units.
The vertical dashed blue line represents zero shift.
See \sect{sp_si4}.
} 
\label{sp_ee}
\end{figure*}
%<<<<<<<<<<<<<<<<<<<<<<<<<<<<<<<<<<<<<<<<<<<<<<<<<<<<<<<<<<<<<<<<<<<<<<<<<<<<<<

%>>>>>>>>>>>>>>>>>>>>>>>>>>>>>>>>>>>>>>>>>>>>>>>>>>>>>>>>>>>>>>>>>>>>>>>>>>>>>>
\begin{figure*}[ht]
\centering {\includegraphics[width=15cm]{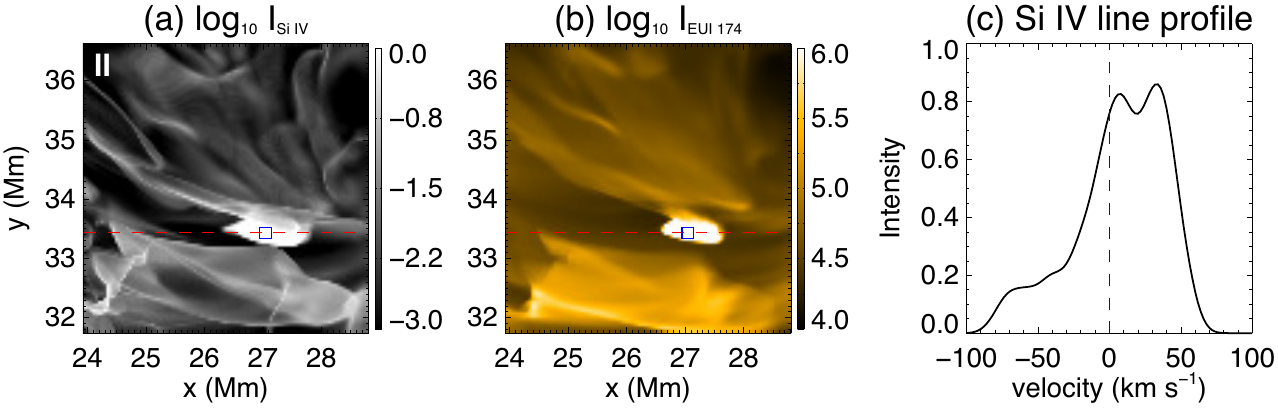}} 
\caption{
Similar to \fig{sp_ee} but for Event 2.
See \sect{sp_si4}.
} 
\label{sp_cf}
\end{figure*}
%<<<<<<<<<<<<<<<<<<<<<<<<<<<<<<<<<<<<<<<<<<<<<<<<<<<<<<<<<<<<<<<<<<<<<<<<<<<<<<

%>>>>>>>>>>>>>>>>>>>>>>>>>>>>>>>>>>>>>>>>>>>>>>>>>>>>>>>>>>>>>>>>>>>>>>>>>>>>>>
\begin{figure*}[ht]
\centering {\includegraphics[width=15cm]{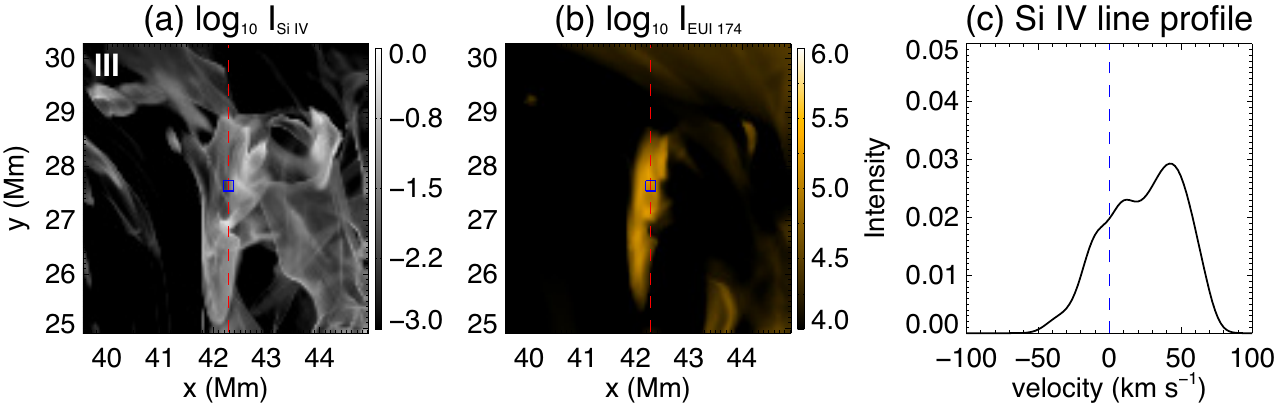}} 
\caption{
Similar to \fig{sp_ee} but for Event 3.
See \sect{sp_si4}.
} 
\label{sp_case3}
\end{figure*}
%<<<<<<<<<<<<<<<<<<<<<<<<<<<<<<<<<<<<<<<<<<<<<<<<<<<<<<<<<<<<<<<<<<<<<<<<<<<<<<

In this section, we present the synthesized transition region and narrowband coronal images as well as the Si~{\sc{iv}} line profiles within the events analyzed in this study as an overview.
The first example is shown in \fig{sp_ee}.
This event exhibits as a compact transient brightening in the Si~{\sc{iv}} intensity map.
It also reveals faint emission enhancement in EUI 174 {\AA} passband.
The blue box in \fig{sp_ee} (a) corresponds to one pixel in the Si~{\sc{iv}} line profile maps with a pixel size of 240 km.
The corresponding Si~{\sc{iv}} line profile shown in \fig{sp_ee}(c) clearly exhibits two peaks, one at the blue wing with a Doppler shift of $\sim$50 km s$^{-1}$ and the other at the red wing with a shift of $\sim$20 km s$^{-1}$.
The two peaks of the line profile represent the bi-directional flows along the vertical direction within the blue box.
We refer this case as Event 1.

The second example presented in \fig{sp_cf} exhibits great emission enhancement in both the Si~{\sc{iv}} line and EUI 174 {\AA} image.
This example was also identified as a transient coronal brightening analyzed in \citet{Chen2021} (Fig. 3 therein), and here we present another snapshot which corresponds to the earliest time when the case is visible as a brightening in the synthesized EUI 174 {\AA} images.
The line profile within the example shown in \fig{sp_cf}(c) strongly deviate from Gaussian.
The profiles at other snapshots also show wing enhancement but not as obvious as the one shown in \fig{sp_cf}(c) comparing to emission near the line center.
There is one peak at the red wing with a Doppler shift of $\sim$35 km s$^{-1}$ except for the peak near the rest wavelength,
and the strong blue wing enhancement extends up to $\sim$70 km s$^{-1}$.
We have reported that this events is generated by magnetic reconnection between two adjacent bundles of field lines in \citet{Chen2021}.
The reconnection outflows in and around the reconnection site are associated with the emission enhancement at wings, and as a result, the Si~{\sc{iv}} line exhibits a multi-peak broadened profile.
We refer this case as Event 2.

The third example presented in \fig{sp_case3} also reveals enhanced emission in both Si~{\sc{iv}} line and EUI 174 {\AA} passband.
It is worth mentioning that the intensity enhancement in 174 {\AA} passband is not strong enough to be identified as a brightening in \citet{Chen2021}.
The Si~{\sc{iv}} line profile identified as an explosive events was 60 s before the snapshot shown in the \fig{sp_case3} at the same location, and the profile exhibits double peaks similar to the one shown in \fig{sp_ee}.
In \fig{sp_case3}, we chose the time when the Si~{\sc{iv}} line and EUI 174 {\AA} intensities reach maximum,
and the Si~{\sc{iv}} profile is still non-Gaussian with red wing enhancement and strong line broadening.
We refer this case as Event 3.
In this paper, we focus on these three examples for a detailed case study.
{The overview of the other events is shown in \fig{extra_events}.}

%>>>>>>>>>>>>>>>>>>>>>>>>>>>>>>>>>>>>>>>>>>>>>>>>>>>>>>>>>>>>>>>>>>>>>>>>>>>>>>
\begin{figure*}
\centering
\includegraphics[width=15cm]{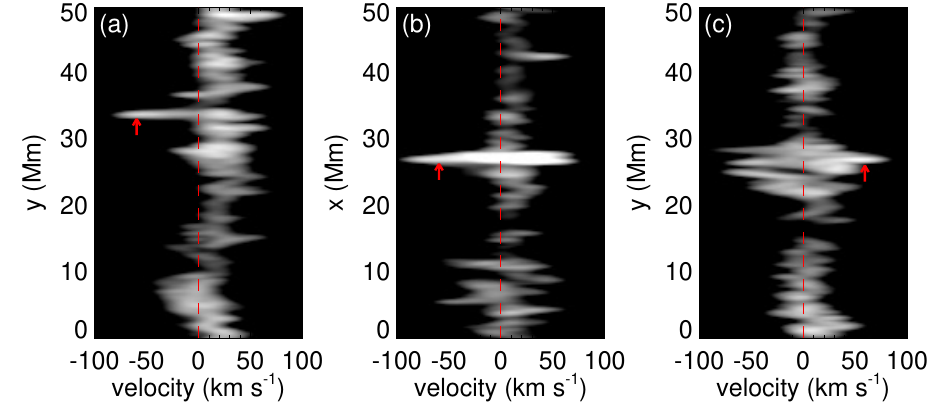}
\caption{{(a) Spectral image of the Si~{\sc{iv}} line across Event 1 along the red dashed line shown in \fig{sp_ee}(a).
The vertical dashed red line represents zero shift.
The red arrow points to the location of Event 1.
(b) Similar to (a) but for Event 2 shown in \fig{sp_cf}.
(c) Similar to (a) but for Event 3 shown in \fig{sp_case3}.
See \sect{sp_si4}.
}}\label{spectroimages}
\end{figure*}
%<<<<<<<<<<<<<<<<<<<<<<<<<<<<<<<<<<<<<<<<<<<<<<<<<<<<<<<<<<<<<<<<<<<<<<<<<<<<<<

%>>>>>>>>>>>>>>>>>>>>>>>>>>>>>>>>>>>>>>>>>>>>>>>>>>>>>>>>>>>>>>>>>>>>>>>>>>>>>>
\begin{figure*}
\centering
\includegraphics[width=15cm]{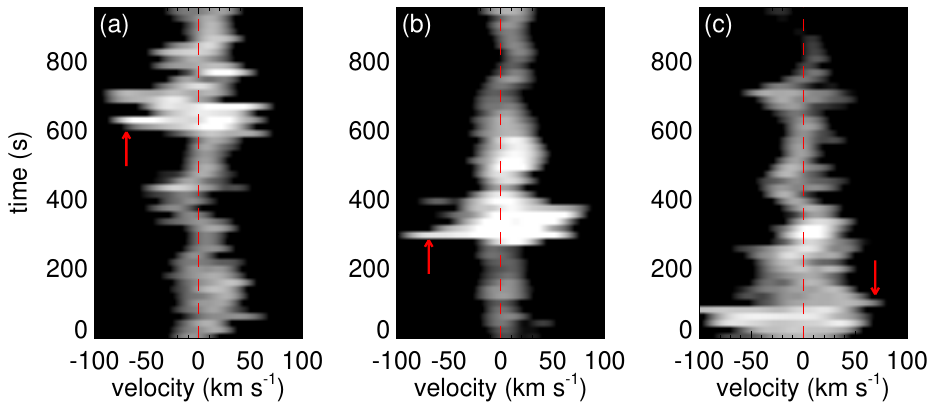}
\caption{
{(a) Temporal evolution of the Si~{\sc{iv}} line profile within the blue box at Event 1 shown in \fig{sp_ee}(a).
The vertical dashed red line represents zero shift.
The red arrow points to the moment shown in \fig{sp_ee}.
(b) Similar to (a) but for Event 2 shown in \fig{sp_cf}.
(c) Similar to (a) but for Event 3 shown in \fig{sp_case3}.
See \sect{sp_si4}.
}
}\label{spectrotime}
\end{figure*}
%<<<<<<<<<<<<<<<<<<<<<<<<<<<<<<<<<<<<<<<<<<<<<<<<<<<<<<<<<<<<<<<<<<<<<<<<<<<<<<

{In addition, we obtained the spectral images of the Si~{\sc{iv}} line across the three events by using slits that were 50 Mm in length, positioned along the red dashed lines shown in \figs{sp_ee}--\ref{sp_case3}.
These spectral images are presented in \fig{spectroimages}.
The three events exhibit clear wing enhancements that are distinct from other quiet Sun regions, similar to observations of explosive events described by \citet{Brueckner1983}.
Furthermore, we present the temporal evolution of the Si~{\sc{iv}} line profiles at the locations of the three events in \fig{spectrotime}.
The wing enhancements of these events were maintained for approximately two minutes, similar to the findings reported by \citet{Dere1989}.
}

%------------------------------------------------------------------------------
\subsection{Evolution of magnetic field structure}\label{S:mag}
%------------------------------------------------------------------------------

%>>>>>>>>>>>>>>>>>>>>>>>>>>>>>>>>>>>>>>>>>>>>>>>>>>>>>>>>>>>>>>>>>>>>>>>>>>>>>>
\begin{figure*}[ht]
\centering {\includegraphics[width=14cm]{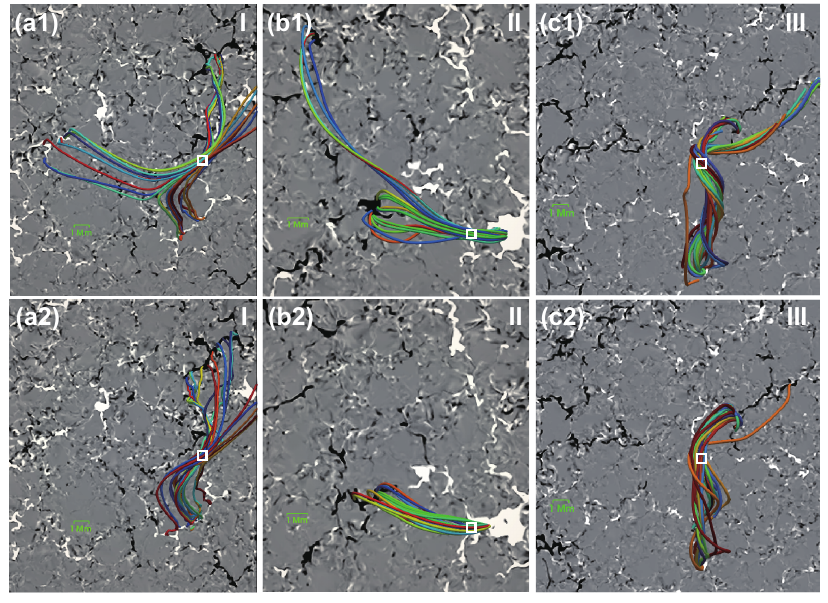}} 
\caption{
Magnetic field structure evolution. 
(a1) The background shows {a zoomed-in view of} the vertical magnetic field in the photosphere, and the magnetogram saturates at $\pm$350 G.
The curves represent magnetic field lines passing through Event 1.
The corresponding time is the same as \fig{sp_ee}.
(a2) Similar to (a1) but for the magnetogram and magnetic field lines one minute later.
{The white boxes indicate the location of the event.}
(b1-b2) Similar to (a1-a2) but for Event 2.
The time difference of the two panels is 9 minutes.
Reproduced from \citet{Chen2021}.
(c1-c2) Similar to (a1-a2) but for Event 3.
{The field of view for all panels represents a small part of the overall calculation domain.}
See \sect{S:mag}.
} 
\label{bline_ee}
\end{figure*}
%<<<<<<<<<<<<<<<<<<<<<<<<<<<<<<<<<<<<<<<<<<<<<<<<<<<<<<<<<<<<<<<<<<<<<<<<<<<<<<

Since the plasma $\beta$ is considerably less than one around all three events, they are dominated by changes in the magnetic field.
To understand their generation mechanisms, we investigated the temporal evolution of the magnetic field structure in and around these events.
For Event 1, we first selected 20 seed points in and around the region with enhanced Si~{\sc{iv}} intensity when the Si~{\sc{iv}} intensity reaches its peak.
Then we traced the field lines from the seed points in both directions as shown in \fig{bline_ee}(a1).
In order to follow the evolution of the magnetic field structure, we traced the field lines from the same seed points in different snapshots.
We also present the field lines when the explosive event disappears, i.e., one minute after the Si~{\sc{iv}} intensity reaches its peak, in \fig{bline_ee}(a2).
We find that there are two bundles of magnetic field lines interacting in and around Event 1.
When the event disappears, the magnetic field structure changes and only the field lines along the $y$-direction remain.
In other words, the event is associated with magnetic reconnection between two bundles of field lines.
The magnetic field structures during and after the occurrence of Event 2 are presented in \fig{bline_ee}(b1-b2),
and these two panels are reproduced from \citet{Chen2021}.
There are two bundles of field lines forking when Event 2 occurs, and the event is located at the region where the two field line bundles diverge.
After the event disappears, only the field line bundle along the $x$-direction remains.
Thus, both Events 1 \& 2 are triggered by the interaction between two field line bundles.
It is worth mentioning that such magnetic field topology works for most coronal brightenings studied in \citet{Chen2021}.

We also present the magnetic field structure in and around Event 3 in \fig{bline_ee}(c1),
and this event occurs in a highly twisted flux rope.
When the Si~{\sc{iv}} brightening disappears, the field lines are highly relaxed as shown in \fig{bline_ee}(c2).
It is possible that magnetic reconnection between field lines within the flux rope contributes to the relaxation of the flux rope and triggers the explosive event. 
The evolution of the magnetic field structure is similar to one case studied in \citet{Chen2021}, in which the coronal brightening is also associated with an untwisting flux rope.
{It is worth mentioning that the changes in the zoomed-in view of magnetograms are mainly the evolution of internetwork magnetic field patches, while the locations of the magnetic concentrations that form network patterns remain stable during the 16-minute period.}

To summarize, the three events are all associated with changes in magnetic field topology.
Events 1 \& 2 are related to the interaction between two bundles of field lines, while Event 3 occurs in an untwisting flux rope, 
and the magnetic field environments for the explosive events are similar to the brightest coronal brightenings in the model.
Thus, magnetic reconnection with small angles between field line bundles or within the twisted flux rope may trigger explosive events and/or coronal brightenings.
However, some events are heated above 1 MK, while others barely reach typical coronal temperatures.
So it is necessary to understand why similar magnetic field topology result in different thermal responses in different events.

%------------------------------------------------------------------------------
\subsection{Thermal responses}\label{S:slices}
%------------------------------------------------------------------------------

%>>>>>>>>>>>>>>>>>>>>>>>>>>>>>>>>>>>>>>>>>>>>>>>>>>>>>>>>>>>>>>>>>>>>>>>>>>>>>>
\begin{figure*}[ht]
\centering {\includegraphics[width=14cm]{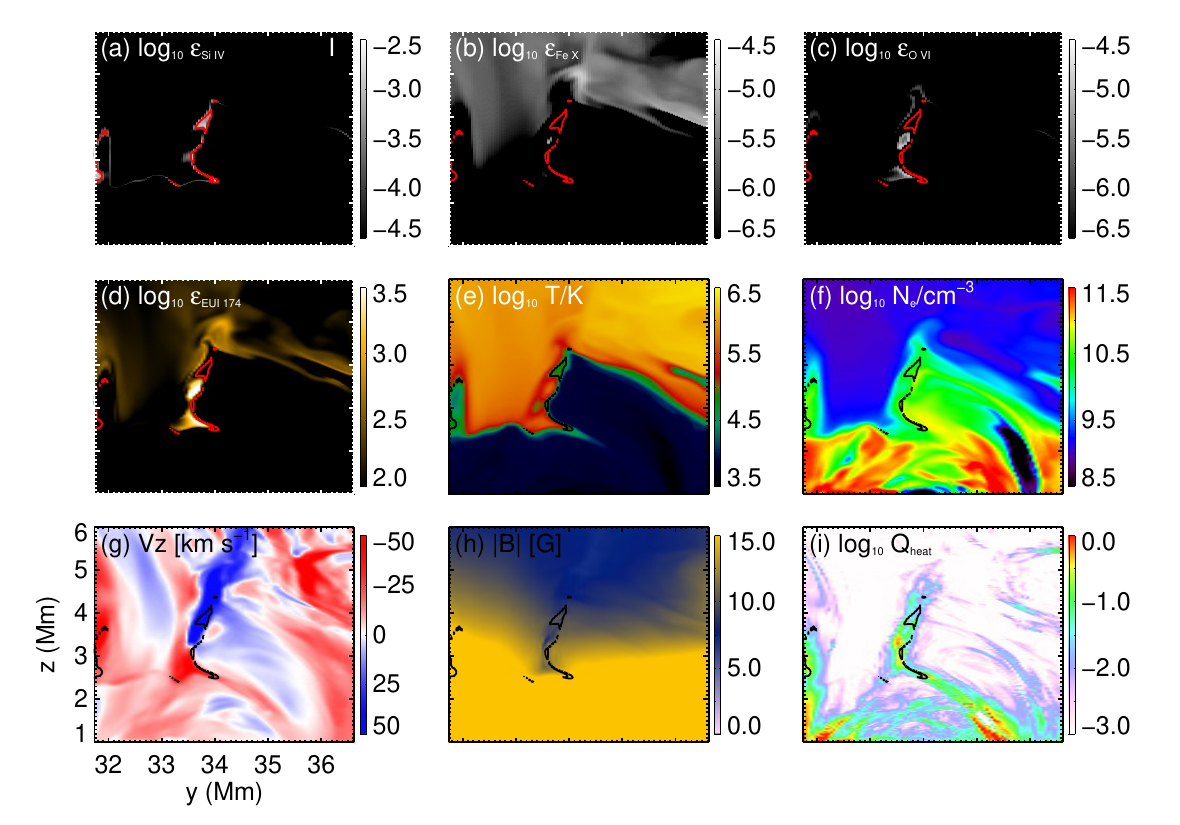}} 
\caption{
Vertical cuts of (a) Si~{\sc{iv}} emission, (b) Fe~{\sc{x}} emission, (c) O~{\sc{vi}} emission, (d) EUI 174 {\AA} emission, (e) temperature, (f) electron number density, (g) vertical velocity, (h) magnetic field strength, and (i) heating rates (sum of viscous and joule heating), respectively, along the red dashed line through Event 1 shown in \fig{sp_ee}.
The emissions are shown in arbitrary units.
The contours outline the regions with enhanced Si~{\sc{iv}} emission.
See \sect{S:slices}.
} 
\label{cut_ee}
\end{figure*}
%<<<<<<<<<<<<<<<<<<<<<<<<<<<<<<<<<<<<<<<<<<<<<<<<<<<<<<<<<<<<<<<<<<<<<<<<<<<<<<

%>>>>>>>>>>>>>>>>>>>>>>>>>>>>>>>>>>>>>>>>>>>>>>>>>>>>>>>>>>>>>>>>>>>>>>>>>>>>>>
\begin{figure*}[ht]
\centering {\includegraphics[width=14cm]{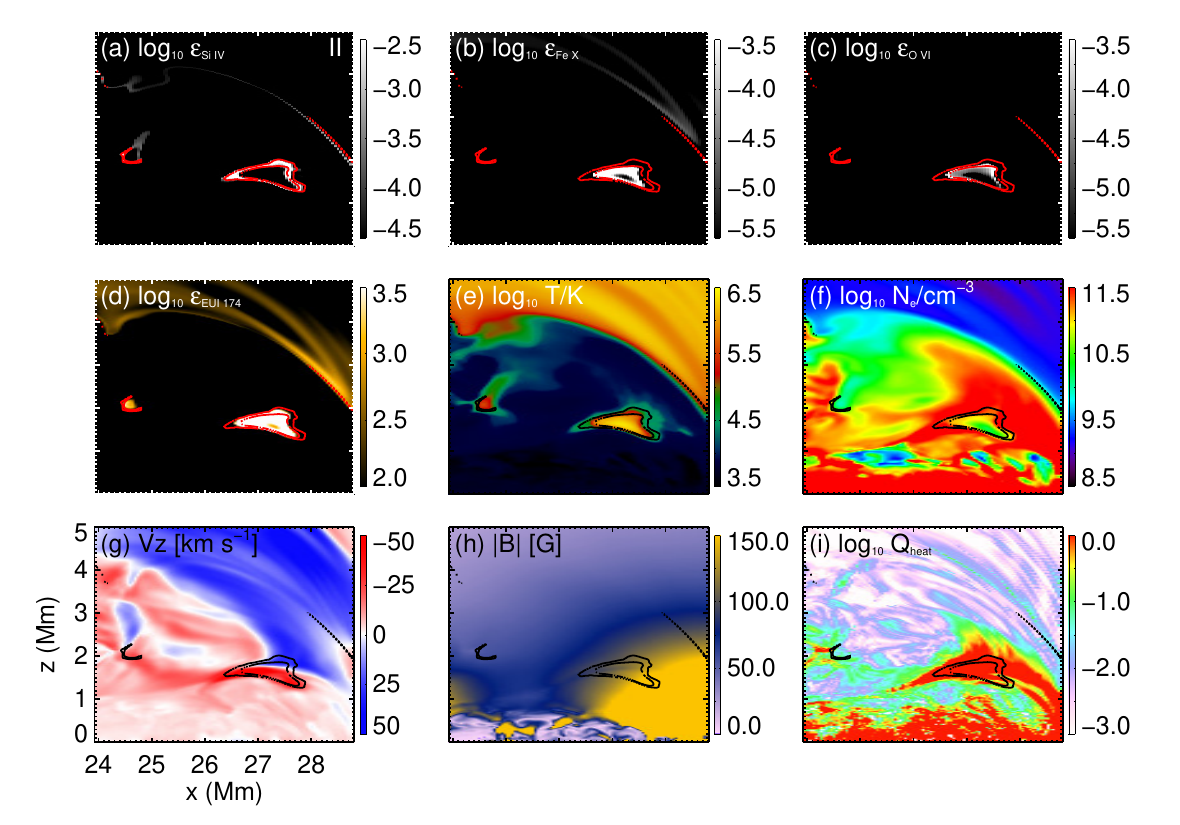}} 
\caption{
Similar to \fig{cut_ee} but along the red dashed line through Event 2 shown in \fig{sp_cf}.
See \sect{S:slices}.
} 
\label{cut_cf}
\end{figure*}
%<<<<<<<<<<<<<<<<<<<<<<<<<<<<<<<<<<<<<<<<<<<<<<<<<<<<<<<<<<<<<<<<<<<<<<<<<<<<<<

%>>>>>>>>>>>>>>>>>>>>>>>>>>>>>>>>>>>>>>>>>>>>>>>>>>>>>>>>>>>>>>>>>>>>>>>>>>>>>>
\begin{figure*}[ht]
\centering {\includegraphics[width=14cm]{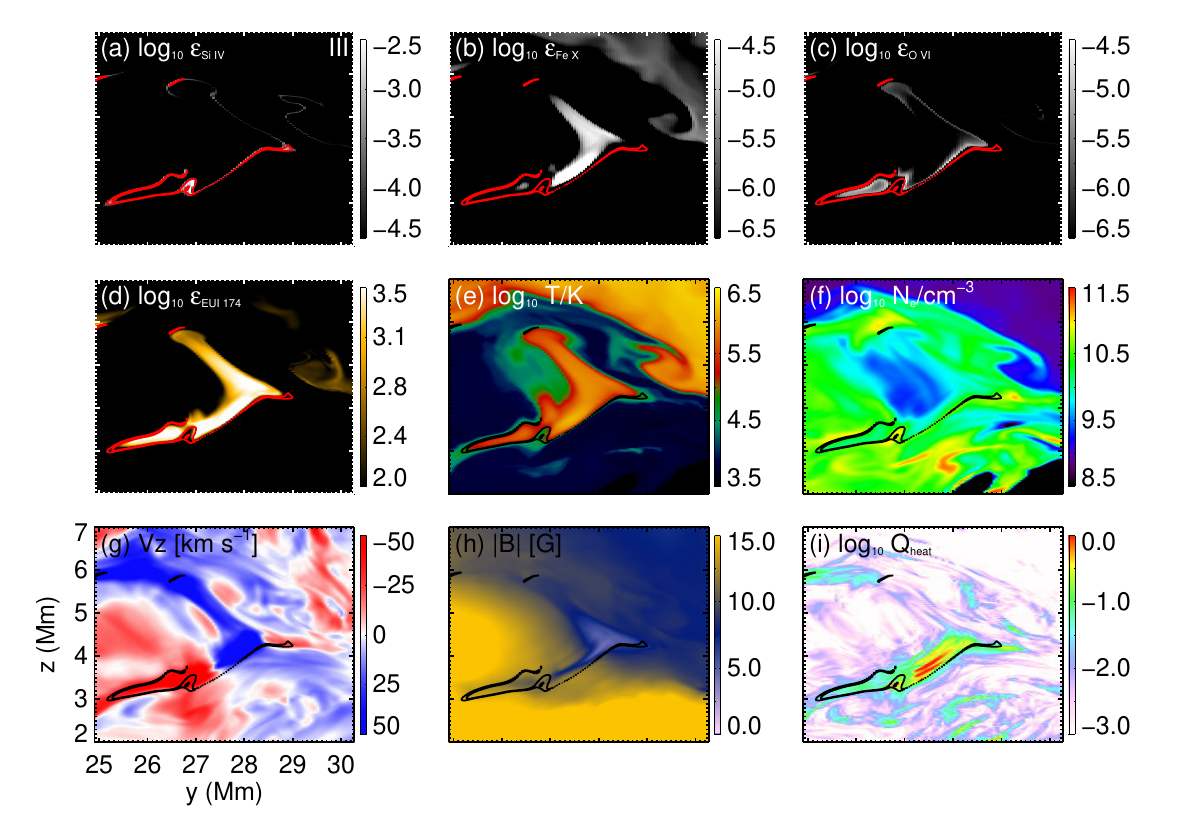}} 
\caption{
Similar to \fig{cut_ee} but along the red dashed line through Event 3 shown in \fig{sp_case3}.
See \sect{S:slices}.
} 
\label{cut_case3}
\end{figure*}
%<<<<<<<<<<<<<<<<<<<<<<<<<<<<<<<<<<<<<<<<<<<<<<<<<<<<<<<<<<<<<<<<<<<<<<<<<<<<<<

To examine the thermal properties of the three examples, we display the vertical cuts of various variables through the events .
As shown in \fig{cut_ee}(a), Event 1 is located between 2.5--4 Mm above the photosphere.
{The vertical velocity map shown in \fig{cut_ee}(g) clearly exhibits signatures of the bi-directional outflows,
which correspond to the two peaks in the Si~{\sc{iv}} line profile shown in \fig{sp_ee}(c).
The decrease in the magnetic field strength and enhancement of the heating rates (sum of viscous and joule heating) around the event (see \fig{cut_ee}(h--i)) imply that {magnetic reconnection takes place},
which is consistent with the magnetic field evolution shown in \fig{bline_ee}.}
%\textcolor{blue}{The authors refer to “reconnection” in the model, but the spatial scale of the simulation is too coarse to resolve reconnection. More details should be given on the process by which bidirectional flows are produced in the model. Are they perhaps produced simply by the plasma being squashed by the dynamics of the field?}
%\todo{Damien, can you write something regarding this comment?}
%
%\DFP{Does this belong here, or back in section 2?}
%\DFP{Due to the limited simulation resolution, resistivity and viscosity in the modeled corona are too high, which will effect the scales on which energy can be deposited.
%
%The model will capture the large scale dissipation of magnetic energy, however the deposition on small scales will not be correctly modeled.
%
%For this reason we study the sum of the resistive and viscous heating.
%}
%
The event also exhibits a little enhanced EUI 174 {\AA} emission next to the Si~{\sc{iv}} brightening.
Considering EUI 174 {\AA} emission has contribution from both the Fe~{\sc{x}} 174 {\AA} line (hot component) and the O~{\sc{vi}} 173 {\AA} line (cool component),
we also present the emission of the two lines in \fig{cut_ee}(b--c).
There is almost no Fe~{\sc{x}} emission within but above the event,
and the emission patterns of EUI 174 {\AA} passband and the O~{\sc{vi}} line overlap significantly.
The temperature within the regions with enhanced EUI 174 {\AA} emission mostly reaches up to 10$^{5.8}$ K.
Thus, this event is dominated by the cool component.

The vertical cuts for Event 2 are presented in \fig{cut_cf}.
This event occurs at the heights of 1.5--2 Mm, and the regions with enhanced Si~{\sc{iv}} emission are surround the EUI 174 {\AA} brightening.
The transition region and coronal emissions are mainly from the shell and core of the brightening, respectively.
Similar to Event 1, there is a decrease in magnetic field strength within the event and the heating rates are enhanced around it,
which are signatures of magnetic reconnection.
The deviation from Gaussian of the Si~{\sc{iv}} line profile stems from the upflows and downflows around the reconnection site.
The temperature of this event reaches above 2 MK, and EUI 174 {\AA} emission is dominated by the Fe~{\sc{x}} line.
In other words, this event is dominated by the hot component.
Compared to Event 1, the number density within the EUI 174 {\AA} brightening is roughly the same (10$^{10}$ cm$^{-3}$).
However, the magnetic field strength around Event 2 is $\sim$150 G, which is about one order of magnitude higher than that surrounding Event 1 ($\sim$15 G).
As a result, the heating rates are much higher in Event 2.
Thus, the temperature difference between Event 1 \& 2 results from the difference in the amount of the converted magnetic energy.

\fig{cut_case3} shows vertical cuts through Event 3.
The vicinity of the reconnection region also reveals a decrease in magnetic field strength, increase in heating rates, and clear bi-directional flow patterns.
The height of this event ranges from 3--5 Mm, and the region with enhanced Si~{\sc{iv}} emission surrounds the EUI 174 {\AA} brightening.
Similar to Event 2, EUI 174 {\AA} emission is dominated by Fe~{\sc{x}}, and the O~{\sc{vi}} emission is mainly located at the edge of the coronal brightening.
In this snapshot, the temperatures within the event reach above 1 MK.
The surrounding magnetic field strength is $\sim$15 G, similar to that of Event 1.
Moreover, the heating rates as well as the converted magnetic energy of Events 1 \& 3 are comparable.
Nevertheless, the density is 10$^{9.5-10}$ cm$^{-3}$, which is much lower than that in Event 1.
Thus, the temperature discrepancy is mainly due to the density difference in and around the reconnection region.

%==============================================================================
\section{Discussions \label{S:diss}}
%==============================================================================

In this study, we first investigated the magnetic field topology around the explosive events.
Two events are located in the region where two bundles of field lines interact, and one example is associated with an untwisting flux rope.
The magnetic field structures greatly change after the explosive events disappear, implying that the events are located at the reconnection region.
The enhancement in current density and heating rates and the decrease in magnetic field strength in all the examples also indicate that magnetic reconnection occurs.
Magnetic reconnection between the field line bundles or within the twisted flux rope triggers the bi-directional flow patterns, which result in the wing enhancement of the Si~{\sc{iv}} line profiles.
\citet{Chen2021} investigated the magnetic field environment of the strongest coronal brightenings in the same model
and found that the energetic coronal events are mostly triggered by component reconnection between the different field line bundles.
Such rearrangement of the tangled magnetic field and associated heating has also been found within cool transition region loops \citep{Li2014}, hot active region loops \citep{Reale2019}, and post-flare loops \citep{Parenti2010}.
In addition, there is one example related to the relaxation of a highly twisted flux rope, which may explain the small-scale transient coronal brightenings accompanied by cool plasma structure \citep{Panesar2021}.
In other words, different types of reconnection events, e.g., transient coronal brightenings and explosive events, in our self-consistent quiet-Sun model may share the common magnetic field environments, i.e., component reconnection.

The scenario of component reconnection, either between different field line bundles or within a flux rope, triggering the explosive events is different from magnetic field setups of most 2D reconnection models of explosive events, in which explosive events are often generated through anti-parallel reconnection \citep[e.g.,][]{Roussev2001a,Innes2015,Peter2019}.
The magnetic field setups in the 2D models are inspired by the connection between explosive events and magnetic field changes in the photosphere, i.e., flux emergence and cancellation \citep[e.g.,][]{Chae1998,Muglach2008}.
In a 3D model of an ephemeral emerging flux region with a higher level of photospheric magnetic field activities, \citet{Tiwari2022} found that some small-scale coronal brightenings are associated with magnetic reconnection between emerging and emerged/preexisting flux.
However, the magnetic field in the photosphere is generated and maintained by the small-scale dynamo in our model, and hence our model represents the very quiet Sun without large-scale flux emergence or preexisting fields.
The lower level of photospheric magnetic activities in our model might explain the reason that the selected explosive events and coronal brightenings in our model occur at coronal heights and are not directly related to the magnetic evolution in the photosphere.
At least, our model can explain explosive events and coronal brightenings without signatures of magnetic field evolution in the photosphere \citep{Muglach2008,Panesar2021}.

Although sharing similar magnetic field environments, the temperature of these events can be quite different.
Events 2 \& 3 are heated above 1 MK, while Event 1 barely reaches typical coronal temperatures.
In principle, the temperature depends on density and the amount of converted energy, or heating rates.
Compared to Event 1, the magnetic field strength and hence heating rates are higher in Event 2 and the density is comparable, while the density is much lower in Event 3 and the heating rates are similar.
Thus, the peak temperature of reconnection events is constrained by both heating rates and density.
Using a series of X-type neutral point reconnection models, \citet{Peter2019} suggested that the maximum temperature in the vicinity of reconnection events and plasma $\beta$ follow a power-law distribution.
We will perform statistical analysis of explosive events in our model to examine whether the relationship between peak temperatures and plasma $\beta$ also follows a power-law distribution as a subsequent study.

%>>>>>>>>>>>>>>>>>>>>>>>>>>>>>>>>>>>>>>>>>>>>>>>>>>>>>>>>>>>>>>>>>>>>>>>>>>>>>>
\begin{figure*}[ht]
\centering {\includegraphics[width=\textwidth]{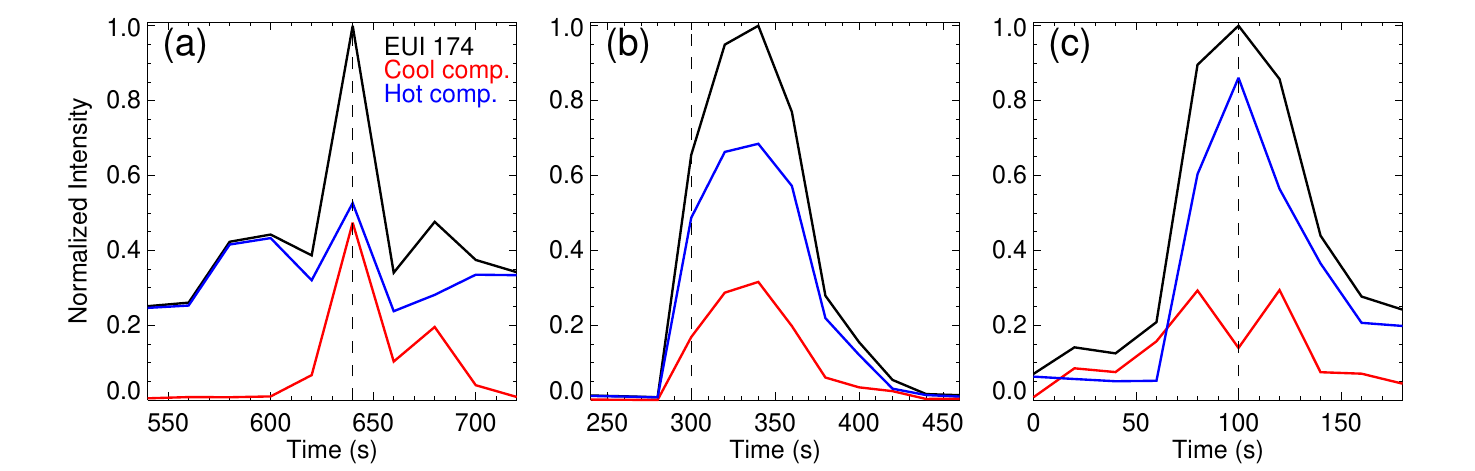}} 
\caption{
(a) Intensity variations of EUI 174 {\AA} passband in Event 1. The black solid curves represent the intensity variations of synthesized EUI 174 {\AA} emission within the blue box shown in \fig{sp_ee}.
Blue and red curves represent the cool and hot components, respectively.
(b-c) Similar to (a) but for Event 2 \& 3 shown in \figs{sp_cf} and \ref{sp_case3}, respectively.
For each panel, the three light curves are normalized by the maximum value of the black curve.
The vertical black dashed lines represent the time of the snapshots shown in \figs{sp_ee}, \ref{sp_cf} and \ref{sp_case3}, respectively.
See \sect{S:diss}.
} 
\label{lc_174}
\end{figure*}
%<<<<<<<<<<<<<<<<<<<<<<<<<<<<<<<<<<<<<<<<<<<<<<<<<<<<<<<<<<<<<<<<<<<<<<<<<<<<<<

%
%Furthermore, it is suggested that the counterparts of explosive events in narrowband coronal images may be mainly due to the contribution from the transition region line within the spectral windows \citep[e.g.,][]{Winebarger2002}.
%
In order to quantitatively investigate the contribution of plasma with transition region and coronal temperatures, respectively, we artificially divided the EUI 174 {\AA} emission into two parts: cool and hot components ($I_{174}^{h}$ and $I_{174}^{c}$ calculated in \sect{S:model}).
The intensity variations of the EUI 174 {\AA} passband and its two components of the three events are shown in \fig{lc_174}.
Event 1 barely reaches coronal temperatures, and the variation of EUI 174 {\AA} intensity is dominated by the cool component.
The hot component is higher than the cool component,
but the hot component is mainly from the coronal emission above the event (see \fig{cut_ee}) and exhibits as a background emission fluctuation.
Events 2 \& 3 are heated up to typical coronal temperatures, and their EUI 174 {\AA} emission are dominated by the hot component.

%>>>>>>>>>>>>>>>>>>>>>>>>>>>>>>>>>>>>>>>>>>>>>>>>>>>>>>>>>>>>>>>>>>>>>>>>>>>>>>
\begin{figure*}[ht]
\centering {\includegraphics[width=\textwidth]{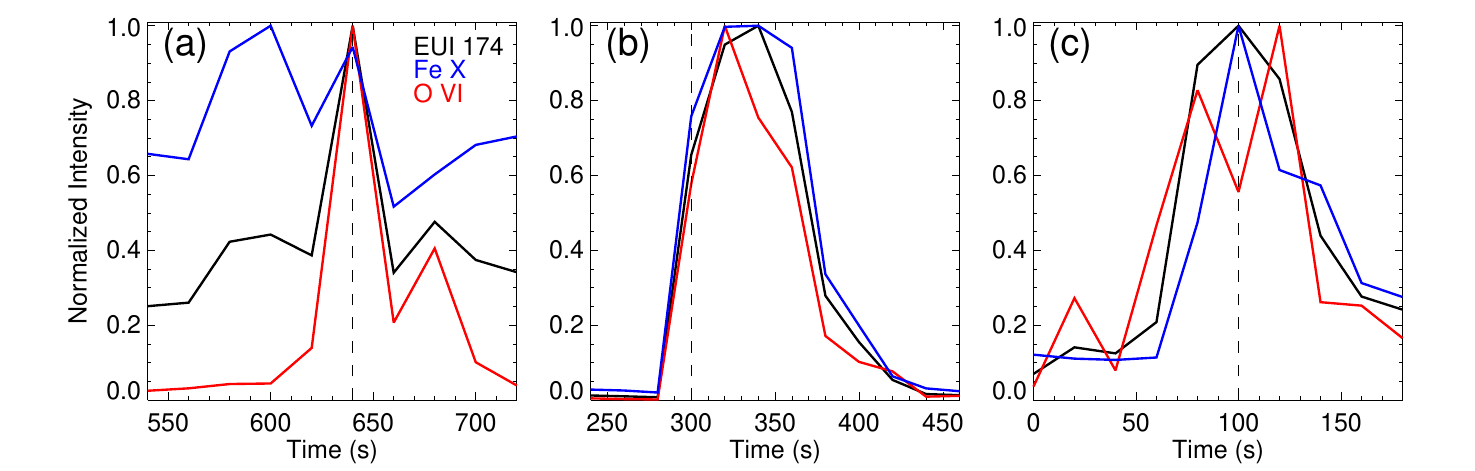}} 
\caption{
Similar to \fig{lc_aia} but for the intensity variations of the synthesized Fe~{\sc{vi}} (blue) and O~{\sc{vi}} (red) lines.
The light curves of EUI 174 {\AA} passband are also presented as black solid lines.
See \sect{S:diss}.
} 
\label{lc_sp}
\end{figure*}
%<<<<<<<<<<<<<<<<<<<<<<<<<<<<<<<<<<<<<<<<<<<<<<<<<<<<<<<<<<<<<<<<<<<<<<<<<<<<<<

In addition, the intensity variations of the Fe~{\sc{x}} and O~{\sc{vi}} are shown in \fig{lc_sp}.
We used the O~{\sc{vi}} 173 {\AA} line because of its contribution to EUI 174 {\AA} passband.
Explosive events can also be identified from the O~{\sc{vi}} 1032 {\AA} line \citep[e.g.,][]{Teriaca2004}, and \citet{Huang2023} investigated spectral observations including the 1032 {\AA} line of small-scale coronal brightenings taken by SPICE.
We synthesized the intensity maps of the O~{\sc{vi}} 1032 {\AA} line for several snapshots and compared them to the 173 {\AA} line intensity maps.
The intensity patterns of these two lines are almost the same, although the intensity ratio changes a lot.
It is because the formation temperatures of the two O~{\sc{vi}} lines are almost the same while their line ratio is sensitive to the temperature.
Moreover, the intensity variation of the 1032 {\AA} line exhibits a similar tendency to the 173 {\AA} line.
Thus we chose the 173 {\AA} line for the following analyses.
The light curves of the O~{\sc{vi}} and Fe~{\sc{x}} line roughly follow the cool and hot components of EUI 174 {\AA} passband shown in \fig{lc_174}, respectively.
For Event 1, the O~{\sc{vi}} line exhibits significant emission enhancement when the EUI 174 {\AA} emission peaks, but the Fe~{\sc{x}} line does not.
It is because the intensity variation in this event is dominated by the cool component and the O~{\sc{vi}} line, and the Fe~{\sc{x}} emission is mainly from the coronal background above the event.
For Event 2, intensity variations of the Fe~{\sc{x}} and O~{\sc{vi}} lines are both similar to that of EUI 174 {\AA}.
At the first snapshot when Event 2 occurs, its temperatures have reached above 2 MK at the core and decrease to 10$^{4.5}$ K at the edge.
Thus the hot and cool components increase simultaneously at the beginning.
During the magnetic reconnection process, the pressure near the reconnection sites is greatly enhanced, and plasma therein expands.
As a result, density drops and the emissions from different temperature ranges decrease together.
The light curve of the Fe~{\sc{x}} line in Event 3 also follows the tendency in that of EUI 174 {\AA}.
Interestingly, intensity variation of the O~{\sc{vi}} line in Event 3 shows double peaks, and the Fe~{\sc{x}} line reaches its peak in between the separated peaks of the O~{\sc{vi}} line.
It is consistent with one event investigated in \citet{Huang2023}.
However, Event 3 cannot fully reproduce the features of the case in their observations.
They found that the light curve of the C~{\sc{iii}} (formation temperature of $10^{4.85}$ K) line also exhibits double peaks, and the O~{\sc{vi}} and Ne~{\sc{viii}} lines reach their peaks in between.
It was explained as the heating and cooling of the event.
We examined the light curve of the Si~{\sc{iv}} line in Event 3, and it only shows one peak at $t=40$ s.
After examining vertical cuts for this event at different snapshots, we found that the average temperature within the EUI 174 {\AA} brightening gradually increases from $\leq10^{4.0}$ K at $t=0$~s to $\sim10^{6.0}$ K at $t=100$~s and slightly decreases at $t=120$~s.
So the plasma heating and subsequent cooling can explain the Fe~{\sc{x}} intensity peak in between the two peaks of the O~{\sc{vi}} line.
Then the density quickly drops since $t=140$~s due to fast untwisting of the flux rope.
It results in rapid intensity decrease in both the Fe~{\sc{x}} and O~{\sc{vi}} lines because the emissivity is proportional to $n_e^2$.
In other words, the intensity variations in Events 2 \& 3 are first dominated by temperature changes and then density variations.
Our analyses are based on examining of vertical cuts, and corks may be needed to be introduced into similar models to draw a more comprehensive and accurate picture of the temporal evolution of temperature and density and their responses in emissions \citep{Druett2022}.

%>>>>>>>>>>>>>>>>>>>>>>>>>>>>>>>>>>>>>>>>>>>>>>>>>>>>>>>>>>>>>>>>>>>>>>>>>>>>>>
\begin{figure*}[ht]
\centering {\includegraphics[width=\textwidth]{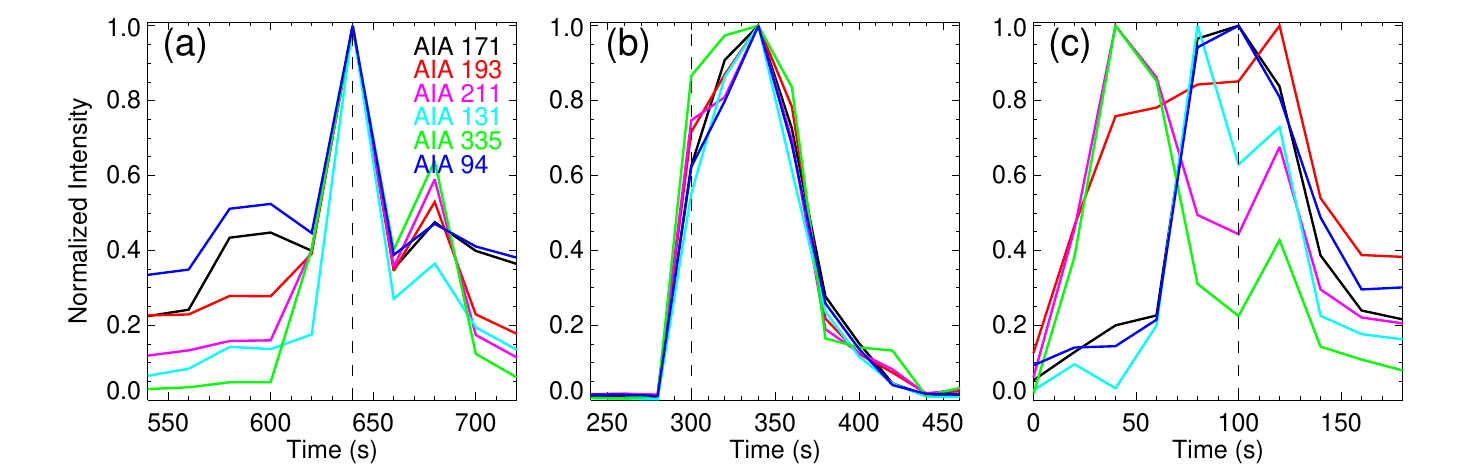}} 
\caption{
Similar to \fig{lc_174} but for the light curves of the synthesized AIA images at different passbands.
Each curve is normalized by its maximum value.
See \sect{S:diss}.
} 
\label{lc_aia}
\end{figure*}
%<<<<<<<<<<<<<<<<<<<<<<<<<<<<<<<<<<<<<<<<<<<<<<<<<<<<<<<<<<<<<<<<<<<<<<<<<<<<<<

It is straightforward to tell whether the events are heated to typical coronal temperatures from spectral observations of coronal emission lines, but it is very challenging to isolate the contribution of hot and cool components in coronal imaging observations.
All AIA coronal channels have contribution from the transition region lines \citep{ODwyer2010,DelZanna2011}, and forward modeling with 3D MHD models indicates that the cool component cannot be neglected \citep{Martinez-Sykora2011}.
\citet{Winebarger2013} suggested that some small-scale active region loops never reach 1 MK as their intensity in different AIA passbands peak at roughly the same time.
Furthermore, \citet{Dolliou2023} performed a statistical time lag analysis with intensity variations of some transient small-scale coronal brightenings in EUI 174 {\AA} and AIA coronal passbands.
They found that the time lags are negligible and suggested that those events are dominated by cool components.
To compare with their findings, we calculated the AIA emissions in different passbands and presented their light curves within the three events investigated in this study in \fig{lc_aia}.
Intensity variations in AIA passbands match very well for both Events 1 \& 2, while Event 1 is dominated by the cool component and Event 2 is multi-thermal and dominated by the hot component.
The cool and hot components of EUI 174 {\AA} emission show similar variations in Event 2, and it is also valid for all AIA coronal passbands.
The cadence of our model is 20 s, so it is also possible that the current temporal resolution of our model may not be sufficient to resolve the heating procedures and hence the time lags between intensities in different passbands.
The correlation among light curves of different passbands in Event 3 is weak.
The peaks near $t=40$ s in 193 and 335 {\AA} passbands are mainly contributed from the transition region lines with formation temperatures $\leq10^{5.5}$ K.
The heating processes are more gentle than in Event 1, and the temperature gradually increases, resulting in various responses in different passbands.
Thus, our analyses suggest that the co-temporal intensity variations in all AIA passbands have two possible explanations:
the event is dominated by the cool component or covers a wide temperature range with heating on short time scales.

%==============================================================================
\section{Conclusions\label{S:conclusions}}
%==============================================================================

In this study, we isolated three explosive events in a 3D self-consistent MHD model of a quiet Sun region.
The Si~{\sc{iv}} line profiles within the events are highly deviated from Gaussian, indicating bi-directional flows triggered by magnetic reconnection.
{They appear along network lanes and last for approximately two minutes.}
Two events are located at the interacting regions of field line bundles, and one example is associated with a highly twisted flux rope.
Their surrounding magnetic field structures are similar to those of transient coronal brightenings studied in \citet{Chen2021}.
In addition, the magnetic field topologies greatly change after the events disappear and the heating rates in and around the events are strong.
Thus, explosive events in our model are manifestations of component reconnection.

Although explosive events may share a similar magnetic field environment, their peak temperatures can be very different.
The plasma within Event 1, {like most events,} barely reaches above 1 MK, while Events 2 \& 3 both achieve typical coronal temperatures.
By examining the vertical cuts of various parameters through the events, we found that temperature differences can result from a different amount of converted magnetic energy, i.e., heating rates, or local plasma density.

Furthermore, we investigated the light curves of AIA coronal passbands.
Both Events 1 \& 2 exhibit simultaneous intensity variations in all AIA passbands, while they are dominated by the cool and hot components, respectively.
So the negligible time lags among AIA passbands cannot tell whether the event is dominated by plasma at transition region temperatures.
It is also possible that the heating processes are too fast to be resolved in observations.
Moreover, the Fe~{\sc{x}} emission greatly increases within the events dominated by the hot component and exhibits as background fluctuations within event dominated by the cool component.
Thus, spectral observations are necessary to determine the thermal properties, and future space missions such as Multi-slit Solar Explorer \citep[MUSE;][]{MUSE} and EUV High-Throughput Spectroscopic Telescope \citep[EUVST;][]{EUVST} are important to unveil their detailed thermal properties and heating processes.

%==============================================================================
%==============================================================================
%==============================================================================
%==============================================================================
\begin{acknowledgements}
{We thank the anonymous referee for constructive comments.}
The work of Y.C. and D.P. was supported by the Deutsches Zentrum f{\"u}r Luft und Raumfahrt (DLR; German Aerospace Center) by grant DLR-FKZ 50OU2201.
{We thank Dr. Luca Teriaca for helpful discussion.}
This research was supported by the International Space Science Institute (ISSI) in Bern, through ISSI International Team project {\#}23-586 (Novel Insights Into Bursts, Bombs, and Brightenings in the Solar Atmosphere from Solar Orbiter).
\end{acknowledgements}

\bibliography{refs}{}

\begin{thebibliography}{63}
\expandafter\ifx\csname natexlab\endcsname\relax\def\natexlab#1{#1}\fi

\bibitem[{{Antolin} {et~al.}(2021){Antolin}, {Pagano}, {Testa}, {Petralia}, \&
  {Reale}}]{Antolin2021}
{Antolin}, P., {Pagano}, P., {Testa}, P., {Petralia}, A., \& {Reale}, F. 2021,
  Nature Astronomy, 5, 54

\bibitem[{{Aschwanden} {et~al.}(2000){Aschwanden}, {Tarbell}, {Nightingale},
  {Schrijver}, {Title}, {Kankelborg}, {Martens}, \& {Warren}}]{Aschwanden2000}
{Aschwanden}, M.~J., {Tarbell}, T.~D., {Nightingale}, R.~W., {et~al.} 2000,
  \apj, 535, 1047

\bibitem[{{Berghmans} {et~al.}(2021){Berghmans}, {Auch{\`e}re}, {Long},
  {Soubri{\'e}}, {Mierla}, {Zhukov}, {Sch{\"u}hle}, {Antolin}, {Harra},
  {Parenti}, {Podladchikova}, {Aznar Cuadrado}, {Buchlin}, {Dolla}, {Verbeeck},
  {Gissot}, {Teriaca}, {Haberreiter}, {Katsiyannis}, {Rodriguez}, {Kraaikamp},
  {Smith}, {Stegen}, {Rochus}, {Halain}, {Jacques}, {Thompson}, \&
  {Inhester}}]{Berghmans2021}
{Berghmans}, D., {Auch{\`e}re}, F., {Long}, D.~M., {et~al.} 2021, \aap, 656, L4

\bibitem[{{Berghmans} {et~al.}(1998){Berghmans}, {Clette}, \&
  {Moses}}]{Berghmans1998}
{Berghmans}, D., {Clette}, F., \& {Moses}, D. 1998, \aap, 336, 1039

\bibitem[{{Bose} {et~al.}(2019){Bose}, {Henriques}, {Joshi}, \& {Rouppe van der
  Voort}}]{Bose2019}
{Bose}, S., {Henriques}, V. M.~J., {Joshi}, J., \& {Rouppe van der Voort}, L.
  2019, \aap, 631, L5

\bibitem[{{Brueckner} \& {Bartoe}(1983)}]{Brueckner1983}
{Brueckner}, G.~E. \& {Bartoe}, J. D.~F. 1983, \apj, 272, 329

\bibitem[{{Chae} {et~al.}(1998){Chae}, {Wang}, {Lee}, {Goode}, \&
  {Sch{\"u}hle}}]{Chae1998}
{Chae}, J., {Wang}, H., {Lee}, C.-Y., {Goode}, P.~R., \& {Sch{\"u}hle}, U.
  1998, \apjl, 497, L109

\bibitem[{{Chen} {et~al.}(2020){Chen}, {Zhang}, {De Pontieu}, {Ma}, {Kliem}, \&
  {Priest}}]{Chen2020}
{Chen}, H., {Zhang}, J., {De Pontieu}, B., {et~al.} 2020, \apj, 899, 19

\bibitem[{{Chen} {et~al.}(2017){Chen}, {Zhang}, {Ma}, {Yan}, \&
  {Xue}}]{Chen2017}
{Chen}, H., {Zhang}, J., {Ma}, S., {Yan}, X., \& {Xue}, J. 2017, \apjl, 841,
  L13

\bibitem[{{Chen} {et~al.}(2022){Chen}, {Peter}, {Przybylski}, {Tian}, \&
  {Zhang}}]{Chen2022}
{Chen}, Y., {Peter}, H., {Przybylski}, D., {Tian}, H., \& {Zhang}, J. 2022,
  \aap, 661, A94

\bibitem[{{Chen} {et~al.}(2021){Chen}, {Przybylski}, {Peter}, {Tian},
  {Auch{\`e}re}, \& {Berghmans}}]{Chen2021}
{Chen}, Y., {Przybylski}, D., {Peter}, H., {et~al.} 2021, \aap, 656, L7

\bibitem[{{Chen} {et~al.}(2019){Chen}, {Tian}, {Huang}, {Peter}, \&
  {Samanta}}]{Chen2019}
{Chen}, Y., {Tian}, H., {Huang}, Z., {Peter}, H., \& {Samanta}, T. 2019, \apj,
  873, 79

\bibitem[{{Chitta} {et~al.}(2021){Chitta}, {Peter}, \& {Young}}]{Chitta2021}
{Chitta}, L.~P., {Peter}, H., \& {Young}, P.~R. 2021, \aap, 647, A159

\bibitem[{{De Pontieu} {et~al.}(2022){De Pontieu}, {Testa},
  {Mart{\'\i}nez-Sykora}, {Antolin}, {Karampelas}, {Hansteen}, {Rempel},
  {Cheung}, {Reale}, {Danilovic}, {Pagano}, {Polito}, {De Moortel},
  {N{\'o}brega-Siverio}, {Van Doorsselaere}, {Petralia}, {Asgari-Targhi},
  {Boerner}, {Carlsson}, {Chintzoglou}, {Daw}, {DeLuca}, {Golub}, {Matsumoto},
  {Ugarte-Urra}, {McIntosh}, \& {the MUSE Team}}]{MUSE}
{De Pontieu}, B., {Testa}, P., {Mart{\'\i}nez-Sykora}, J., {et~al.} 2022, \apj,
  926, 52

\bibitem[{{De Pontieu} {et~al.}(2014){De Pontieu}, {Title}, {Lemen}, {Kushner},
  {Akin}, {Allard}, {Berger}, {Boerner}, {Cheung}, {Chou}, {Drake}, {Duncan},
  {Freeland}, {Heyman}, {Hoffman}, {Hurlburt}, {Lindgren}, {Mathur}, {Rehse},
  {Sabolish}, {Seguin}, {Schrijver}, {Tarbell}, {W{\"u}lser}, {Wolfson},
  {Yanari}, {Mudge}, {Nguyen-Phuc}, {Timmons}, {van Bezooijen}, {Weingrod},
  {Brookner}, {Butcher}, {Dougherty}, {Eder}, {Knagenhjelm}, {Larsen},
  {Mansir}, {Phan}, {Boyle}, {Cheimets}, {DeLuca}, {Golub}, {Gates}, {Hertz},
  {McKillop}, {Park}, {Perry}, {Podgorski}, {Reeves}, {Saar}, {Testa}, {Tian},
  {Weber}, {Dunn}, {Eccles}, {Jaeggli}, {Kankelborg}, {Mashburn}, {Pust},
  {Springer}, {Carvalho}, {Kleint}, {Marmie}, {Mazmanian}, {Pereira}, {Sawyer},
  {Strong}, {Worden}, {Carlsson}, {Hansteen}, {Leenaarts}, {Wiesmann},
  {Aloise}, {Chu}, {Bush}, {Scherrer}, {Brekke}, {Martinez-Sykora}, {Lites},
  {McIntosh}, {Uitenbroek}, {Okamoto}, {Gummin}, {Auker}, {Jerram}, {Pool}, \&
  {Waltham}}]{IRIS}
{De Pontieu}, B., {Title}, A.~M., {Lemen}, J.~R., {et~al.} 2014, \solphys, 289,
  2733

\bibitem[{{Del Zanna} {et~al.}(2021){Del Zanna}, {Dere}, {Young}, \&
  {Landi}}]{CHIANTI10}
{Del Zanna}, G., {Dere}, K.~P., {Young}, P.~R., \& {Landi}, E. 2021, \apj, 909,
  38

\bibitem[{{Del Zanna} {et~al.}(2011){Del Zanna}, {O'Dwyer}, \&
  {Mason}}]{DelZanna2011}
{Del Zanna}, G., {O'Dwyer}, B., \& {Mason}, H.~E. 2011, \aap, 535, A46

\bibitem[{{Dere}(1994)}]{Dere1994AdSpR}
{Dere}, K.~P. 1994, Advances in Space Research, 14, 13

\bibitem[{{Dere} {et~al.}(1989){Dere}, {Bartoe}, \& {Brueckner}}]{Dere1989}
{Dere}, K.~P., {Bartoe}, J. D.~F., \& {Brueckner}, G.~E. 1989, \solphys, 123,
  41

\bibitem[{{Dere} {et~al.}(1991){Dere}, {Bartoe}, {Brueckner}, {Ewing}, \&
  {Lund}}]{Dere1991JGR}
{Dere}, K.~P., {Bartoe}, J. D.~F., {Brueckner}, G.~E., {Ewing}, J., \& {Lund},
  P. 1991, \jgr, 96, 9399

\bibitem[{{Dere} {et~al.}(1997){Dere}, {Landi}, {Mason}, {Monsignori Fossi}, \&
  {Young}}]{CHIANTI}
{Dere}, K.~P., {Landi}, E., {Mason}, H.~E., {Monsignori Fossi}, B.~C., \&
  {Young}, P.~R. 1997, \aaps, 125, 149

\bibitem[{{Dolliou} {et~al.}(2023){Dolliou}, {Parenti}, {Auch{\`e}re},
  {Bocchialini}, {Pelouze}, {Antolin}, {Berghmans}, {Harra}, {Long},
  {Sch{\"u}hle}, {Kraaikamp}, {Stegen}, {Verbeeck}, {Gissot}, {Aznar Cuadrado},
  {Buchlin}, {Mierla}, {Teriaca}, \& {Zhukov}}]{Dolliou2023}
{Dolliou}, A., {Parenti}, S., {Auch{\`e}re}, F., {et~al.} 2023, \aap, 671, A64

\bibitem[{{Dolliou} {et~al.}(2024){Dolliou}, {Parenti}, \&
  {Bocchialini}}]{2024A&A...688A..77D}
{Dolliou}, A., {Parenti}, S., \& {Bocchialini}, K. 2024, \aap, 688, A77

\bibitem[{{Druett} {et~al.}(2022){Druett}, {Leenaarts}, {Carlsson}, \&
  {Szydlarski}}]{Druett2022}
{Druett}, M.~K., {Leenaarts}, J., {Carlsson}, M., \& {Szydlarski}, M. 2022,
  \aap, 665, A6

\bibitem[{{Handy} {et~al.}(1999){Handy}, {Acton}, {Kankelborg}, {Wolfson},
  {Akin}, {Bruner}, {Caravalho}, {Catura}, {Chevalier}, {Duncan}, {Edwards},
  {Feinstein}, {Freeland}, {Friedlaender}, {Hoffmann}, {Hurlburt}, {Jurcevich},
  {Katz}, {Kelly}, {Lemen}, {Levay}, {Lindgren}, {Mathur}, {Meyer}, {Morrison},
  {Morrison}, {Nightingale}, {Pope}, {Rehse}, {Schrijver}, {Shine}, {Shing},
  {Strong}, {Tarbell}, {Title}, {Torgerson}, {Golub}, {Bookbinder}, {Caldwell},
  {Cheimets}, {Davis}, {Deluca}, {McMullen}, {Warren}, {Amato}, {Fisher},
  {Maldonado}, \& {Parkinson}}]{TRACE}
{Handy}, B.~N., {Acton}, L.~W., {Kankelborg}, C.~C., {et~al.} 1999, \solphys,
  187, 229

\bibitem[{{Hansteen} {et~al.}(2019){Hansteen}, {Ortiz}, {Archontis},
  {Carlsson}, {Pereira}, \& {Bj{\o}rgen}}]{Hansteen2019}
{Hansteen}, V., {Ortiz}, A., {Archontis}, V., {et~al.} 2019, \aap, 626, A33

\bibitem[{{Hansteen} {et~al.}(2017){Hansteen}, {Archontis}, {Pereira},
  {Carlsson}, {Rouppe van der Voort}, \& {Leenaarts}}]{Hansteen2017}
{Hansteen}, V.~H., {Archontis}, V., {Pereira}, T.~M.~D., {et~al.} 2017, \apj,
  839, 22

\bibitem[{{Huang} {et~al.}(2023){Huang}, {Teriaca}, {Aznar Cuadrado}, {Chitta},
  {Mandal}, {Peter}, {Sch{\"u}hle}, {Solanki}, {Auch{\`e}re}, {Berghmans},
  {Buchlin}, {Carlsson}, {Fludra}, {Fredvik}, {Giunta}, {Grundy}, {Hassler},
  {Parenti}, \& {Plaschke}}]{Huang2023}
{Huang}, Z., {Teriaca}, L., {Aznar Cuadrado}, R., {et~al.} 2023, \aap, 673, A82

\bibitem[{{Innes}(2001)}]{Innes2001}
{Innes}, D.~E. 2001, \aap, 378, 1067

\bibitem[{{Innes} {et~al.}(1997{\natexlab{a}}){Innes}, {Brekke}, {Germerott},
  \& {Wilhelm}}]{Innes1997SoPh}
{Innes}, D.~E., {Brekke}, P., {Germerott}, D., \& {Wilhelm}, K.
  1997{\natexlab{a}}, \solphys, 175, 341

\bibitem[{{Innes} {et~al.}(2015){Innes}, {Guo}, {Huang}, \&
  {Bhattacharjee}}]{Innes2015}
{Innes}, D.~E., {Guo}, L.~J., {Huang}, Y.~M., \& {Bhattacharjee}, A. 2015,
  \apj, 813, 86

\bibitem[{{Innes} {et~al.}(1997{\natexlab{b}}){Innes}, {Inhester}, {Axford}, \&
  {Wilhelm}}]{Innes1997}
{Innes}, D.~E., {Inhester}, B., {Axford}, W.~I., \& {Wilhelm}, K.
  1997{\natexlab{b}}, \nat, 386, 811

\bibitem[{{Innes} \& {T{\'o}th}(1999)}]{Innes1999}
{Innes}, D.~E. \& {T{\'o}th}, G. 1999, \solphys, 185, 127

\bibitem[{{Joshi} \& {Rouppe van der Voort}(2022)}]{Joshi2022}
{Joshi}, J. \& {Rouppe van der Voort}, L. H.~M. 2022, \aap, 664, A72

\bibitem[{{Kahil} {et~al.}(2022){Kahil}, {Hirzberger}, {Solanki}, {Chitta},
  {Peter}, {Auch{\`e}re}, {Sinjan}, {Orozco Su{\'a}rez}, {Albert}, {Albelo
  Jorge}, {Appourchaux}, {Alvarez-Herrero}, {Blanco Rodr{\'\i}guez},
  {Gandorfer}, {Germerott}, {Guerrero}, {Guti{\'e}rrez M{\'a}rquez}, {Kolleck},
  {del Toro Iniesta}, {Volkmer}, {Woch}, {Fiethe}, {G{\'o}mez Cama},
  {P{\'e}rez-Grande}, {Sanchis Kilders}, {Balaguer Jim{\'e}nez}, {Bellot
  Rubio}, {Calchetti}, {Carmona}, {Deutsch}, {Fern{\'a}ndez-Rico},
  {Fern{\'a}ndez-Medina}, {Garc{\'\i}a Parejo}, {Gasent-Blesa}, {Gizon},
  {Grauf}, {Heerlein}, {Lagg}, {Lange}, {L{\'o}pez Jim{\'e}nez}, {Maue},
  {Meller}, {Michalik}, {Moreno Vacas}, {M{\"u}ller}, {Nakai}, {Schmidt},
  {Schou}, {Sch{\"u}hle}, {Staub}, {Strecker}, {Torralbo}, {Valori}, {Aznar
  Cuadrado}, {Teriaca}, {Berghmans}, {Verbeeck}, {Kraaikamp}, \&
  {Gissot}}]{Kahil2022}
{Kahil}, F., {Hirzberger}, J., {Solanki}, S.~K., {et~al.} 2022, \aap, 660, A143

\bibitem[{{Lemen} {et~al.}(2012){Lemen}, {Title}, {Akin}, {Boerner}, {Chou},
  {Drake}, {Duncan}, {Edwards}, {Friedlaender}, {Heyman}, {Hurlburt}, {Katz},
  {Kushner}, {Levay}, {Lindgren}, {Mathur}, {McFeaters}, {Mitchell}, {Rehse},
  {Schrijver}, {Springer}, {Stern}, {Tarbell}, {Wuelser}, {Wolfson}, {Yanari},
  {Bookbinder}, {Cheimets}, {Caldwell}, {Deluca}, {Gates}, {Golub}, {Park},
  {Podgorski}, {Bush}, {Scherrer}, {Gummin}, {Smith}, {Auker}, {Jerram},
  {Pool}, {Soufli}, {Windt}, {Beardsley}, {Clapp}, {Lang}, \& {Waltham}}]{AIA}
{Lemen}, J.~R., {Title}, A.~M., {Akin}, D.~J., {et~al.} 2012, \solphys, 275, 17

\bibitem[{{Li} {et~al.}(2014){Li}, {Peter}, {Chen}, \& {Zhang}}]{Li2014}
{Li}, L.~P., {Peter}, H., {Chen}, F., \& {Zhang}, J. 2014, \aap, 570, A93

\bibitem[{Lloyd(1982)}]{kmeans}
Lloyd, S. 1982, IEEE Transactions on Information Theory, 28, 129

\bibitem[{{Madjarska}(2019)}]{Madjarska2019}
{Madjarska}, M.~S. 2019, Living Reviews in Solar Physics, 16, 2

\bibitem[{{Mart{\'\i}nez-Sykora} {et~al.}(2011){Mart{\'\i}nez-Sykora}, {De
  Pontieu}, {Testa}, \& {Hansteen}}]{Martinez-Sykora2011}
{Mart{\'\i}nez-Sykora}, J., {De Pontieu}, B., {Testa}, P., \& {Hansteen}, V.
  2011, \apj, 743, 23

\bibitem[{{Muglach}(2008)}]{Muglach2008}
{Muglach}, K. 2008, \apj, 687, 1398

\bibitem[{{M{\"u}ller} {et~al.}(2020){M{\"u}ller}, {St. Cyr}, {Zouganelis},
  {Gilbert}, {Marsden}, {Nieves-Chinchilla}, {Antonucci}, {Auch{\`e}re},
  {Berghmans}, {Horbury}, {Howard}, {Krucker}, {Maksimovic}, {Owen}, {Rochus},
  {Rodriguez-Pacheco}, {Romoli}, {Solanki}, {Bruno}, {Carlsson}, {Fludra},
  {Harra}, {Hassler}, {Livi}, {Louarn}, {Peter}, {Sch{\"u}hle}, {Teriaca}, {del
  Toro Iniesta}, {Wimmer-Schweingruber}, {Marsch}, {Velli}, {De Groof},
  {Walsh}, \& {Williams}}]{SolO}
{M{\"u}ller}, D., {St. Cyr}, O.~C., {Zouganelis}, I., {et~al.} 2020, \aap, 642,
  A1

\bibitem[{{Ni} {et~al.}(2021){Ni}, {Chen}, {Peter}, {Tian}, \& {Lin}}]{Ni2021}
{Ni}, L., {Chen}, Y., {Peter}, H., {Tian}, H., \& {Lin}, J. 2021, \aap, 646,
  A88

\bibitem[{{O'Dwyer} {et~al.}(2010){O'Dwyer}, {Del Zanna}, {Mason}, {Weber}, \&
  {Tripathi}}]{ODwyer2010}
{O'Dwyer}, B., {Del Zanna}, G., {Mason}, H.~E., {Weber}, M.~A., \& {Tripathi},
  D. 2010, \aap, 521, A21

\bibitem[{{Panesar} {et~al.}(2021){Panesar}, {Tiwari}, {Berghmans}, {Cheung},
  {M{\"u}ller}, {Auchere}, \& {Zhukov}}]{Panesar2021}
{Panesar}, N.~K., {Tiwari}, S.~K., {Berghmans}, D., {et~al.} 2021, \apjl, 921,
  L20

\bibitem[{{Parenti} {et~al.}(2010){Parenti}, {Reale}, \&
  {Reeves}}]{Parenti2010}
{Parenti}, S., {Reale}, F., \& {Reeves}, K.~K. 2010, \aap, 517, A41

\bibitem[{{Peter} {et~al.}(2019){Peter}, {Huang}, {Chitta}, \&
  {Young}}]{Peter2019}
{Peter}, H., {Huang}, Y.~M., {Chitta}, L.~P., \& {Young}, P.~R. 2019, \aap,
  628, A8

\bibitem[{{Purkhart} \& {Veronig}(2022)}]{Purkhart2022}
{Purkhart}, S. \& {Veronig}, A.~M. 2022, \aap, 661, A149

\bibitem[{{Reale} {et~al.}(2019){Reale}, {Testa}, {Petralia}, \&
  {Graham}}]{Reale2019}
{Reale}, F., {Testa}, P., {Petralia}, A., \& {Graham}, D.~R. 2019, \apj, 882, 7

\bibitem[{{Rempel}(2017)}]{MURaM2017}
{Rempel}, M. 2017, \apj, 834, 10

\bibitem[{{Rochus} {et~al.}(2020){Rochus}, {Auch{\`e}re}, {Berghmans}, {Harra},
  {Schmutz}, {Sch{\"u}hle}, {Addison}, {Appourchaux}, {Aznar Cuadrado},
  {Baker}, {Barbay}, {Bates}, {BenMoussa}, {Bergmann}, {Beurthe}, {Borgo},
  {Bonte}, {Bouzit}, {Bradley}, {B{\"u}chel}, {Buchlin}, {B{\"u}chner},
  {Cab{\'e}}, {Cadiergues}, {Chaigneau}, {Chares}, {Choque Cortez}, {Coker},
  {Condamin}, {Coumar}, {Curdt}, {Cutler}, {Davies}, {Davison}, {Defise}, {Del
  Zanna}, {Delmotte}, {Delouille}, {Dolla}, {Dumesnil}, {D{\"u}rig}, {Enge},
  {Fran{\c{c}}ois}, {Fourmond}, {Gillis}, {Giordanengo}, {Gissot}, {Green},
  {Guerreiro}, {Guilbaud}, {Gyo}, {Haberreiter}, {Hafiz}, {Hailey}, {Halain},
  {Hansotte}, {Hecquet}, {Heerlein}, {Hellin}, {Hemsley}, {Hermans}, {Hervier},
  {Hochedez}, {Houbrechts}, {Ihsan}, {Jacques}, {J{\'e}r{\^o}me}, {Jones},
  {Kahle}, {Kennedy}, {Klaproth}, {Kolleck}, {Koller}, {Kotsialos},
  {Kraaikamp}, {Langer}, {Lawrenson}, {Le Clech'}, {Lenaerts}, {Liebecq},
  {Linder}, {Long}, {Mampaey}, {Markiewicz-Innes}, {Marquet}, {Marsch},
  {Matthews}, {Mazy}, {Mazzoli}, {Meining}, {Meltchakov}, {Mercier}, {Meyer},
  {Monecke}, {Monfort}, {Morinaud}, {Moron}, {Mountney}, {M{\"u}ller},
  {Nicula}, {Parenti}, {Peter}, {Pfiffner}, {Philippon}, {Phillips},
  {Plesseria}, {Pylyser}, {Rabecki}, {Ravet-Krill}, {Rebellato}, {Renotte},
  {Rodriguez}, {Roose}, {Rosin}, {Rossi}, {Roth}, {Rouesnel}, {Roulliay},
  {Rousseau}, {Ruane}, {Scanlan}, {Schlatter}, {Seaton}, {Silliman}, {Smit},
  {Smith}, {Solanki}, {Spescha}, {Spencer}, {Stegen}, {Stockman}, {Szwec},
  {Tamiatto}, {Tandy}, {Teriaca}, {Theobald}, {Tychon}, {van Driel-Gesztelyi},
  {Verbeeck}, {Vial}, {Werner}, {West}, {Westwood}, {Wiegelmann}, {Willis},
  {Winter}, {Zerr}, {Zhang}, \& {Zhukov}}]{EUI}
{Rochus}, P., {Auch{\`e}re}, F., {Berghmans}, D., {et~al.} 2020, \aap, 642, A8

\bibitem[{{Roussev} {et~al.}(2001{\natexlab{a}}){Roussev}, {Doyle},
  {Galsgaard}, \& {Erd{\'e}lyi}}]{Roussev2001c}
{Roussev}, I., {Doyle}, J.~G., {Galsgaard}, K., \& {Erd{\'e}lyi}, R.
  2001{\natexlab{a}}, \aap, 380, 719

\bibitem[{{Roussev} {et~al.}(2001{\natexlab{b}}){Roussev}, {Galsgaard},
  {Erd{\'e}lyi}, \& {Doyle}}]{Roussev2001a}
{Roussev}, I., {Galsgaard}, K., {Erd{\'e}lyi}, R., \& {Doyle}, J.~G.
  2001{\natexlab{b}}, \aap, 370, 298

\bibitem[{{Shimizu} {et~al.}(2019){Shimizu}, {Imada}, {Kawate}, {Ichimoto},
  {Suematsu}, {Hara}, {Katsukawa}, {Kubo}, {Toriumi}, {Watanabe}, {Yokoyama},
  {Korendyke}, {Warren}, {Tarbell}, {De Pontieu}, {Teriaca}, {Sch{\"u}hle},
  {Solanki}, {Harra}, {Matthews}, {Fludra}, {Auch{\`e}re}, {Andretta},
  {Naletto}, \& {Zhukov}}]{EUVST}
{Shimizu}, T., {Imada}, S., {Kawate}, T., {et~al.} 2019, in Society of
  Photo-Optical Instrumentation Engineers (SPIE) Conference Series, Vol. 11118,
  UV, X-Ray, and Gamma-Ray Space Instrumentation for Astronomy XXI, ed. O.~H.
  {Siegmund}, 1111807

\bibitem[{{Skan} {et~al.}(2023){Skan}, {Danilovic}, {Leenaarts}, {Calvo}, \&
  {Rempel}}]{Skan2023}
{Skan}, M., {Danilovic}, S., {Leenaarts}, J., {Calvo}, F., \& {Rempel}, M.
  2023, \aap, 672, A47

\bibitem[{{SPICE Consortium} {et~al.}(2020){SPICE Consortium}, {Anderson},
  {Appourchaux}, {Auch{\`e}re}, {Aznar Cuadrado}, {Barbay}, {Baudin},
  {Beardsley}, {Bocchialini}, {Borgo}, {Bruzzi}, {Buchlin}, {Burton},
  {B{\"u}chel}, {Caldwell}, {Caminade}, {Carlsson}, {Curdt}, {Davenne},
  {Davila}, {Deforest}, {Del Zanna}, {Drummond}, {Dubau}, {Dumesnil}, {Dunn},
  {Eccleston}, {Fludra}, {Fredvik}, {Gabriel}, {Giunta}, {Gottwald}, {Griffin},
  {Grundy}, {Guest}, {Gyo}, {Haberreiter}, {Hansteen}, {Harrison}, {Hassler},
  {Haugan}, {Howe}, {Janvier}, {Klein}, {Koller}, {Kucera}, {Kouliche},
  {Marsch}, {Marshall}, {Marshall}, {Matthews}, {McQuirk}, {Meining},
  {Mercier}, {Morris}, {Morse}, {Munro}, {Parenti}, {Pastor-Santos}, {Peter},
  {Pfiffner}, {Phelan}, {Philippon}, {Richards}, {Rogers}, {Sawyer},
  {Schlatter}, {Schmutz}, {Sch{\"u}hle}, {Shaughnessy}, {Sidher}, {Solanki},
  {Speight}, {Spescha}, {Szwec}, {Tamiatto}, {Teriaca}, {Thompson}, {Tosh},
  {Tustain}, {Vial}, {Walls}, {Waltham}, {Wimmer-Schweingruber}, {Woodward},
  {Young}, {de Groof}, {Pacros}, {Williams}, \& {M{\"u}ller}}]{SPICE}
{SPICE Consortium}, {Anderson}, M., {Appourchaux}, T., {et~al.} 2020, \aap,
  642, A14

\bibitem[{{Teriaca} {et~al.}(2004){Teriaca}, {Banerjee}, {Falchi}, {Doyle}, \&
  {Madjarska}}]{Teriaca2004}
{Teriaca}, L., {Banerjee}, D., {Falchi}, A., {Doyle}, J.~G., \& {Madjarska},
  M.~S. 2004, \aap, 427, 1065

\bibitem[{{Teriaca} {et~al.}(2002){Teriaca}, {Madjarska}, \&
  {Doyle}}]{Teriaca2002}
{Teriaca}, L., {Madjarska}, M.~S., \& {Doyle}, J.~G. 2002, \aap, 392, 309

\bibitem[{{Tian} {et~al.}(2014){Tian}, {DeLuca}, {Cranmer}, {De Pontieu},
  {Peter}, {Mart{\'\i}nez-Sykora}, {Golub}, {McKillop}, {Reeves}, {Miralles},
  {McCauley}, {Saar}, {Testa}, {Weber}, {Murphy}, {Lemen}, {Title}, {Boerner},
  {Hurlburt}, {Tarbell}, {Wuelser}, {Kleint}, {Kankelborg}, {Jaeggli},
  {Carlsson}, {Hansteen}, \& {McIntosh}}]{Tian2014Sci}
{Tian}, H., {DeLuca}, E.~E., {Cranmer}, S.~R., {et~al.} 2014, Science, 346,
  1255711

\bibitem[{{Tiwari} {et~al.}(2022){Tiwari}, {Hansteen}, {De Pontieu}, {Panesar},
  \& {Berghmans}}]{Tiwari2022}
{Tiwari}, S.~K., {Hansteen}, V.~H., {De Pontieu}, B., {Panesar}, N.~K., \&
  {Berghmans}, D. 2022, \apj, 929, 103

\bibitem[{{V{\"o}gler} {et~al.}(2005){V{\"o}gler}, {Shelyag}, {Sch{\"u}ssler},
  {Cattaneo}, {Emonet}, \& {Linde}}]{MURaM}
{V{\"o}gler}, A., {Shelyag}, S., {Sch{\"u}ssler}, M., {et~al.} 2005, \aap, 429,
  335

\bibitem[{{Winebarger} {et~al.}(2002){Winebarger}, {Updike}, \&
  {Reeves}}]{Winebarger2002}
{Winebarger}, A.~R., {Updike}, A.~C., \& {Reeves}, K.~K. 2002, \apjl, 570, L105

\bibitem[{{Winebarger} {et~al.}(2013){Winebarger}, {Walsh}, {Moore}, {De
  Pontieu}, {Hansteen}, {Cirtain}, {Golub}, {Kobayashi}, {Korreck}, {DeForest},
  {Weber}, {Title}, \& {Kuzin}}]{Winebarger2013}
{Winebarger}, A.~R., {Walsh}, R.~W., {Moore}, R., {et~al.} 2013, \apj, 771, 21

\end{thebibliography}
\bibliographystyle{aa}

\begin{appendix}

\section{Additional examples of explosive events}\label{S:appendix.examples}

{In the main text, we presented detailed studies of three examples. Here we provide an overview of other randomly selected events in the same format as in \fig{sp_ee}.}

%>>>>>>>>>>>>>>>>>>>>>>>>>>>>>>>>>>>>>>>>>>>>>>>>>>>>>>>>>>>>>>>>>>>>>>>>>>>>>>
\begin{figure*}
\centering
\includegraphics[width=10cm]{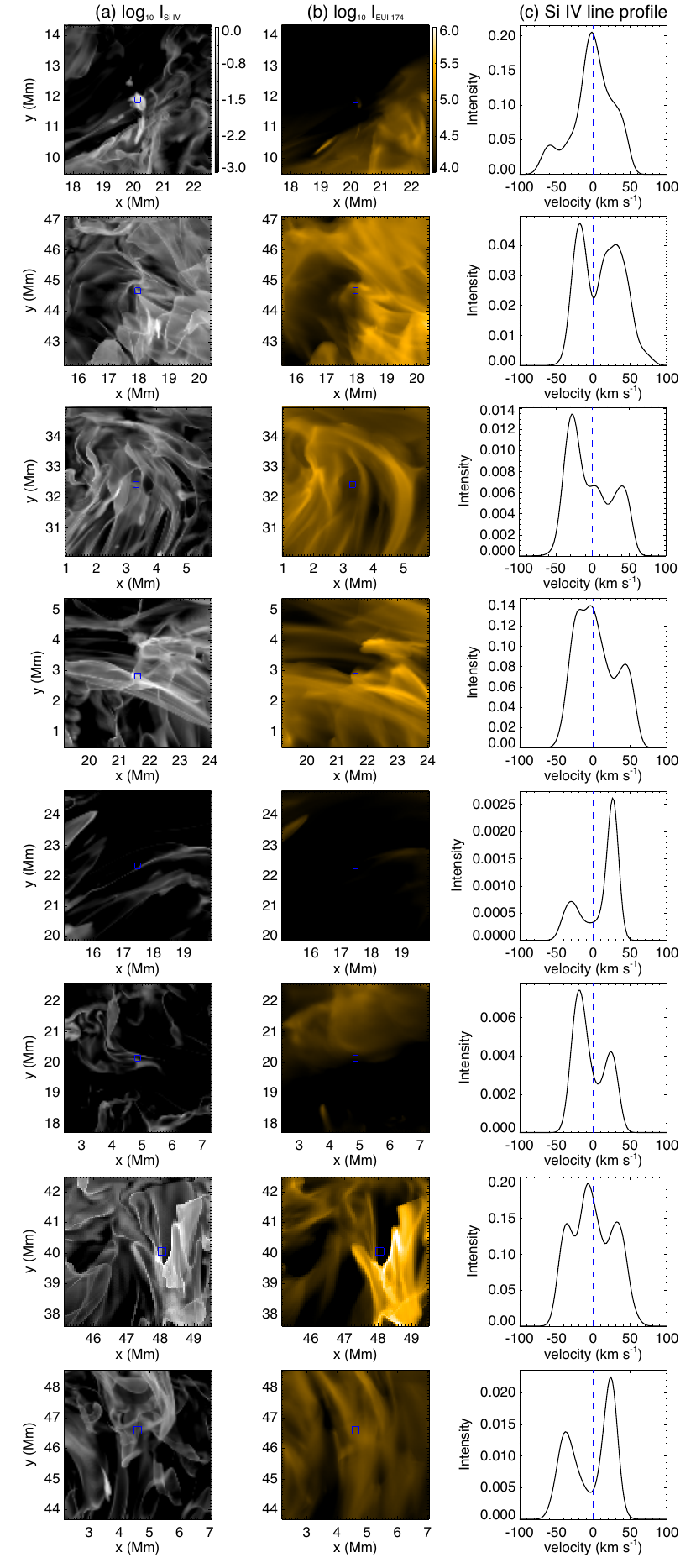}
\caption{{Intensity maps and the Si~{\sc{iv}} line profiles for the other events.
Each row is analogous to \fig{sp_ee} but for different events.
See \sect{S:appendix.examples}.
}
}\label{extra_events}
\end{figure*}
%<<<<<<<<<<<<<<<<<<<<<<<<<<<<<<<<<<<<<<<<<<<<<<<<<<<<<<<<<<<<<<<<<<<<<<<<<<<<<<

\end{appendix}

\end{document}